\documentclass[a4paper,12pt]{article}

\usepackage[T1]{fontenc}
\usepackage[utf8]{inputenc}
\usepackage[british]{babel}
\usepackage{amsmath,amsthm,mathtools}
\usepackage{newtxtext,newtxmath}
\usepackage{bm}
\usepackage{microtype}
\usepackage{geometry}
\usepackage{setspace}
\usepackage[mathlines]{lineno}
\usepackage{graphicx}
\usepackage{tikz}
\usepackage{xcolor}
\usepackage{booktabs,tabularx,threeparttable}
\usepackage{caption}
\usepackage[shortlabels]{enumitem}
\usepackage{natbib}
\usepackage[hidelinks]{hyperref}

\modulolinenumbers[1]

\bibpunct{(}{)}{;}{a}{}{,}

\theoremstyle{plain}
\newtheorem{theorem}{Theorem}[section]
\newtheorem{lemma}[theorem]{Lemma}
\newtheorem{corollary}[theorem]{Corollary}
\newtheorem{definition}[theorem]{Definition}

\theoremstyle{remark}

\numberwithin{equation}{section}
\numberwithin{table}{section}
\numberwithin{figure}{section}

\newcommand{\figref}[1]{Figure~\ref{fig:#1}}
\newcommand{\figsref}[1]{Figures~\ref{fig:#1}}

\newcommand{\secref}[1]{Section~\ref{sec:#1}}

\newcommand{\appref}[1]{Appendix~\ref{app:#1}}

\newcommand{\defref}[1]{Definition~\ref{def:#1}}

\newcommand{\lemref}[1]{Lemma~\ref{lem:#1}}

\newcommand{\thmref}[1]{Theorem~\ref{thm:#1}}

\newcommand{\corref}[1]{Corollary~\ref{cor:#1}}

\newcommand{\tabref}[1]{Table~\ref{tab:#1}}

\newcommand{\cR}{\mathcal{R}}
\newcommand{\cH}{\mathcal{H}}
\newcommand{\cA}{\mathcal{A}}
\newcommand{\Sidak}{\v{S}id\'ak}

\newcommand{\cG}{\mathcal{G}}

\newcommand{\cT}{\mathcal{T}}

\newcommand{\FWER}{\text{FWER}}

\allowdisplaybreaks

\title{Uniformly Improved Adaptive Filtering Procedures for Replicability Testing with FWER Control}
\author{Ninh Tran\\[0.75ex]
  \normalsize School of Mathematics and Statistics, The University of Melbourne}
\date{}

\begin{document}
\maketitle
\begin{center}
\normalsize\textbf{Correspondence:} Ninh Tran, \href{mailto:ninht@student.unimelb.edu.au}{ninht@student.unimelb.edu.au}
\end{center}

\begin{abstract}
Identifying signals that replicate across multiple studies is a fundamental challenge in modern data analysis. Partial conjunction (PC) testing addresses this problem by assessing whether a signal is present in at least a specified number of studies. The AdaFilter framework of Wang et al. (2022, \textit{Annals of Statistics} \textbf{50}, 1890--1909) improves the power of PC testing through adaptive filtering, but its family-wise error rate (FWER)-controlling procedures rely on conservative Bonferroni corrections at two stages: (i) the construction of PC $p$-values and filtering statistics, and (ii) the multiplicity adjustment used to control the FWER. We show that both stages can be improved using the \Sidak\ correction. Under the same independence assumptions as the original AdaFilter framework, our improvements preserve finite-sample FWER control and guarantee at least as many rejections as their Bonferroni counterparts for every realisation of the data. Simulations and a real data analysis demonstrate the practical benefits of these improvements.
\end{abstract}


\section{Introduction}\label{sec:introduction}

Replicability is a fundamental principle of scientific research,
providing evidence that an observed finding is not specific to a
particular study \citep{OpenScienceCollaboration2015,Munafo2017,Nosek2020}.
Concerns about the replicability of published findings have arisen
across several scientific disciplines, with large-scale replication
projects reporting substantial discrepancies between original studies
and their replications \citep{Camerer2018}.
These concerns motivate statistical methods that assess whether a
scientific signal is present across multiple independent studies,
rather than a single study.

Suppose we wish to investigate $m>1$ \textit{features}, such as $m$
different genes in a given tissue.
A feature is considered to exhibit a scientifically
interesting \textit{signal} in a study if its underlying
effect lies outside a prespecified null region.
However, evidence for a signal in one study does not
necessarily imply its presence in other studies.
This may be due to differences
in study populations, designs, and experimental conditions \citep{Nosek2022}.
Meta-analysis provides a natural framework for synthesising
evidence from multiple studies and assessing the replicability
and robustness of scientific signals \citep{Gurevitch2018}.

In this work, we consider a meta-analysis comprising $n>1$
comparable studies, each examining the same set of $m$ features.
Let $[\ell]:=\{1,\dots,\ell\}$ for any positive integer $\ell$.
We let $\theta_{ij}$ denote the effect of feature $i\in[m]$
in study $j\in[n]$, allowing the underlying effects to vary
across studies, and we assume that the null region $\Theta_{0i}$
is common across studies.
The \textit{base} null hypothesis for feature $i$ in study $j$
is defined as
\begin{equation*}
    H_{ij}:\theta_{ij}\in\Theta_{0i},
\end{equation*}
where $\theta_{ij}$ is considered a signal whenever $H_{ij}$ is false.

To formally assess replicability, we consider partial conjunction
(PC) null hypotheses, which allow us to test whether a signal is
present, or \textit{replicated}, in at least a specified number
of studies.
For a prespecified replicability level $u\in[n]$, the PC null for feature $i$ is
\begin{equation*}
    H_i^{u/n}:
    \left|
        \left\{
            j\in[n]:\theta_{ij}\notin\Theta_{0i}
        \right\}
    \right|
    <u.
\end{equation*}
Thus, if $H_i^{u/n}$ is false, then a signal is present in at
least $u$ studies, in which case we say that feature $i$ is
$u/n$ replicated. In particular, $u=1$ corresponds to
the global null hypothesis that $\theta_{ij} \in \Theta_{0i}$
for all $j \in [n]$, whereas $u\geq 2$ allows us to make
claims of replication across multiple studies.
Throughout the remainder of this paper, we focus on
replicability levels $u\in\{2,\dots,n\}$.

Since we consider $m>1$ features, we seek to test the
PC hypotheses $H_1^{u/n},\dots,H_m^{u/n}$
simultaneously.
Let $\cH^{u/n}_0\subseteq[m]$ denote the indices of the
true PC null hypotheses and let $\cR\subseteq[m]$ denote
the set of indices of PC null hypotheses rejected by a
multiple testing procedure.
In this paper, we focus on procedures that control the
family-wise error rate (FWER), defined as
\begin{equation*}
    \FWER(\cR)
    :=
    \Pr\left(
        \cR\cap\cH^{u/n}_0 \neq\varnothing
    \right),
\end{equation*}
at or below a prespecified level $\alpha\in(0,1)$.
In the context of our work, the FWER is the probability
of making at least one false claim of $u/n$ replicability.

The PC testing framework was introduced by \citet{Benjamini2008},
who developed valid $p$-values for testing the PC null
hypothesis $H_i^{u/n}$ by combining the $p$-values
corresponding to the base null hypotheses
$H_{i1},\dots,H_{in}$.
Let $P_{ij}$ denote a valid base $p$-value for testing $H_{ij}$,
and let $P_{i(1)}\leq\dots\leq P_{i(n)}$ denote the ordered
base $p$-values for feature $i$.
A simple construction for testing $H_i^{u/n}$ is the
Bonferroni PC $p$-value, given by
\begin{equation}\label{Bonferroni_PC_p_value}
    S_i=S_i^{u/n}
    := (n-u+1)P_{i(u)} \in [0,n-u+1],
\end{equation}
which satisfies the validity property $\Pr(S_i\leq t)\leq t$ for all $t \in [0,1]$ when $H^{u/n}_i$ 
is true by Theorem 2 of \citet{Benjamini2008}.
The intuition behind this construction is that, when
$H_i^{u/n}$ is true, up to $u-1$ base null hypotheses
may be false and their corresponding $p$-values may be
arbitrarily small. The Bonferroni PC $p$-value therefore
discards the $u-1$ smallest base $p$-values and applies
a Bonferroni correction to the remaining $n-u+1$
$p$-values. Other approaches, including Fisher's combination method
and Simes' correction, can also be used to construct
valid PC $p$-values under appropriate assumptions.

A straightforward approach to controlling FWER is to
apply a standard multiple-testing procedure, such as
the Bonferroni or Holm procedure \citep{Holm1979}, to the $m$ PC
$p$-values $S_1,\dots,S_m$ as defined in \eqref{Bonferroni_PC_p_value}.
For example, the Bonferroni
procedure rejects $H_i^{u/n}$ whenever
$S_i\leq\alpha/m$. Although this approach controls
FWER, it can be severely conservative. The difficulty
arises partly because the PC null hypothesis is highly
composite, allowing different numbers and configurations
of false base null hypotheses.
Consequently, a valid PC $p$-value can be substantially
more conservative than a uniform random variable under
the null \citep{Liang2025}. When such conservative
$p$-values are further subjected to a multiplicity
correction across all $m$ features to control FWER, the resulting
procedure can suffer substantial losses in power,
particularly when $m$ is large and signals are
sparse \citep{Wang2022}.

Several methods have been proposed to improve PC testing,
including more flexible constructions of PC $p$-values
\citep{Wang2019}, methods accommodating dependence in multiple
testing \citep{Bogomolov2023}, conditional approaches
that reduce the conservativeness of PC $p$-values
\citep{Liang2025}, and filtering followed by conditional
PC testing \citep{Dickhaus2026}.

A particularly influential approach to PC testing is AdaFilter,
introduced by \citet{Wang2022}, which improves power by adaptively
filtering out features that are unlikely to be $(u-1)/n$
replicated before testing the remaining features for $u/n$
replicability using the Bonferroni PC $p$-values defined in
\eqref{Bonferroni_PC_p_value}.
This is a natural filtering criterion, since $u/n$
replicability necessarily implies $(u-1)/n$ replicability.
By removing unpromising candidates, AdaFilter reduces the
multiplicity burden of testing for $u/n$ replicability and
can achieve substantial power gains over directly applying
the Bonferroni correction to all $m$ PC $p$-values.
However, since the filtering and testing steps use the same
data, the resulting data-dependent selection may invalidate
FWER control, even when the PC $p$-values are valid.
Crucially, under independence of the base $p$-values, a
\textit{conditional validity} property of the Bonferroni PC
$p$-values, established by \citet{Wang2022}, permits
testing after filtering while maintaining FWER control;
see \secref{adafilter} for details.

Despite these developments, there remains scope to uniformly
improve the power of FWER-controlling procedures.
In particular, the original FWER-controlling procedure
AdaFilter-Bon\footnote{Termed AdaFilter-Bonferroni by
\citet{Wang2022}; we use the abbreviated name
AdaFilter-Bon for consistency with the naming of
the procedures developed in this paper.}
relies on Bonferroni-type corrections at two distinct stages.
First, both its PC $p$-value and filtering statistic are
constructed using the Bonferroni combining function.
Second, its rejection threshold is determined using a
Bonferroni correction to account for multiple testing
among features passing the filter.
While Bonferroni corrections are theoretically justified, they can be
conservative under independence of the base $p$-values.
Under this assumption, the \Sidak\ correction offers a
sharper alternative at both stages.
However, these improvements require theoretical justification,
as the \Sidak-based PC $p$-values must preserve the
conditional validity property that justifies testing
after filtering, while the \Sidak\ multiplicity correction
applied after filtering must still guarantee FWER control.

In this paper, we address both sources of conservativeness
and develop procedures that uniformly improve upon the
original FWER-controlling procedures within the AdaFilter
framework, under the same independence assumptions as
the original framework.
Our main contributions are as follows:

\begin{itemize}
    \item We construct a \Sidak-based PC $p$-value and
a corresponding filtering statistic, both of which
are uniformly no larger than their Bonferroni counterparts.
    We establish that these improved statistics preserve
    the conditional validity property underlying AdaFilter,
    allowing them to replace the original statistics
    without compromising FWER control
    (\secref{uniform_improved_stat}).

    \item Building on these improved statistics, we develop 
    AdaFilter-Sid, which further replaces the Bonferroni
    multiplicity correction across filtered features
    with a sharper \Sidak\ correction.
    We establish finite-sample FWER control and show
    that AdaFilter-Sid uniformly improves upon AdaFilter-Bon,
    rejecting at least as many PC hypotheses for every
    realisation of the data.
    Using the sequential rejection principle of
    \citet{Goeman2010}, we further develop
    AdaFilter-SeqSid, a sequential extension that uniformly
    improves upon both AdaFilter-Sid and the sequential
    extension of AdaFilter-Bon discussed in
    Remark~4.2 of \citet{Wang2022}
(\secref{sidak_fwer}).

\end{itemize}

The remainder of this paper is organised as follows.
In \secref{adafilter}, we review the AdaFilter framework
and its FWER-controlling procedures.
In \secref{uniform_improved_stat}, we introduce the
improved PC $p$-value and filtering statistic, while
\secref{sidak_fwer} develops the proposed single-step
and sequential \Sidak\ procedures.
We investigate their performance through simulations
(\secref{simulations}) and illustrate their application
to real data (\secref{real_data}).
Finally, we conclude with a discussion in
\secref{discussion}.

\section{The AdaFilter Framework}
\label{sec:adafilter}

Recall that, for each $(i,j)\in[m]\times[n]$, $P_{ij}$
denotes a valid $p$-value for testing the base null
hypothesis $H_{ij}$, and that the Bonferroni PC $p$-value
$S_i=S_i^{u/n}$ defined in \eqref{Bonferroni_PC_p_value}
is valid for testing $H_i^{u/n}$.
The AdaFilter framework of \citet{Wang2022} reduces the
multiplicity burden by filtering out features that are
unlikely to be $(u-1)/n$ replicated before testing
the remaining features for $u/n$ replicability.
To this end, for each feature $i\in[m]$ and
$u\in\{2,\dots,n\}$, define the filtering statistic
associated with $S_i$ as
\begin{equation*}
    F_i=(n-u+1)P_{i(u-1)}.
\end{equation*}

The filtering statistic is proportional to the
Bonferroni PC $p$-value for testing $(u-1)/n$
replicability, since
\begin{equation*}
    F_i
    =
    \frac{n-u+1}{n-u+2}S_i^{(u-1)/n}.
\end{equation*}
This relationship motivates the filtering step.
Since $u/n$ replicability implies $(u-1)/n$ replicability,
a small value of $F_i$ indicates preliminary evidence
that feature $i$ is replicated in at least $u-1$ studies,
making it a promising candidate for $u/n$ replicability.

AdaFilter-Bon, originally defined in
\citet[Definition~3.1]{Wang2022} and restated below
in our notation, uses the filtering statistics
$F_1,\dots,F_m$ to identify promising features
and reduce the multiplicity burden when applying
the Bonferroni FWER procedure to the Bonferroni
PC $p$-values.
To quantify the number of features passing the filter
at a candidate threshold $t\in[0,\alpha]$, define
\begin{equation*}
    N(t) :=
    \sum_{i\in[m]}
    \mathbf{1}\{F_i<t\}.
\end{equation*}

\begin{definition}[AdaFilter-Bon]\label{def:adafilter_bon}
For a target FWER level $\alpha\in(0,1)$, define the feasible region
\begin{equation*}
    \cT_{\mathrm{Bon}} := \left\{
        t\in[0,\alpha]:
        t N(t)
        \leq \alpha
    \right\}
\end{equation*}
and the data-dependent threshold $\hat{t}_{\mathrm{Bon}} := \sup \cT_{\mathrm{Bon}}$.
AdaFilter-Bon outputs the rejection set
\begin{equation*}
    \cR_{\mathrm{Bon}}
    :=
    \left\{
        i\in[m]:
        S_i<\hat{t}_{\mathrm{Bon}}
    \right\}.
\end{equation*}
\end{definition}

Define the set of features retained by the filtering
step of AdaFilter-Bon as
$
    \cA
    :=
    \left\{
        i\in[m]:
        F_i<\hat{t}_{\mathrm{Bon}}
    \right\}.
$
Because $F_i\leq S_i$ for each $i\in[m]$,
it follows from \defref{adafilter_bon} that
\begin{equation*}
    \cR_{\mathrm{Bon}}
    =
    \left\{
        i\in\cA:
        S_i<\hat{t}_{\mathrm{Bon}}
    \right\}.
\end{equation*}
Thus, every rejected feature necessarily passes the
filtering step. Moreover, when $\cA\neq\varnothing$,
\begin{equation*}
    \frac{\alpha}{m}
    \leq
    \hat{t}_{\mathrm{Bon}}
    \leq
    \frac{\alpha}{|\cA|} = \frac{\alpha}{N(\hat{t}_{\mathrm{Bon}})}.
\end{equation*}
The lower bound holds because $t=\alpha/m$
always satisfies $tN(t)\leq\alpha$. Hence, AdaFilter-Bon uses a rejection
threshold at least as large as that of the ordinary
Bonferroni procedure, potentially yielding substantial
power gains when relatively few features survive filtering, i.e., $N(\hat{t}_{\mathrm{Bon}})$ is small.

Under mutual independence of all base $p$-values
in $(P_{ij})_{(i,j)\in[m]\times[n]}$,
\citet[Theorem~4.2]{Wang2022} established that
AdaFilter-Bon controls the FWER at level $\alpha$,
i.e.,
\begin{equation*}
    \Pr\left(
        \cR_{\mathrm{Bon}}
        \cap\cH^{u/n}_0 \neq\varnothing
    \right)
    \leq\alpha.
\end{equation*}
A key ingredient in establishing this result is the
\emph{conditional validity} property established
in \citet[Lemma~4.1]{Wang2022}. Specifically,
under independence of the base $p$-values
$P_{i1},\dots,P_{in}$, whenever $H_i^{u/n}$
is true,
\begin{equation}\label{conditional_validity}
    \Pr(S_i<t\mid F_i<t)\leq t
\end{equation}
for every $t\in(0,1]$ such that
$\Pr(F_i<t)>0$. Together with independence across features, this property
allows AdaFilter-Bon to apply a Bonferroni correction
based on the number of features surviving the filtering
step, rather than the total number of features,
while maintaining FWER control.

For FWER control, \citet[Section~S1]{Wang2022}
further strengthened AdaFilter-Bon using the
sequential rejection principle of \citet{Goeman2010},
a general framework that is closely related to
closed testing. The idea is analogous to the
improvement of the Bonferroni procedure by Holm's
procedure: after an initial set of rejections is made,
the rejected hypotheses are removed from the
multiplicity correction, and the AdaFilter-Bon threshold
is recomputed over the remaining hypotheses. This
process is repeated until no further hypotheses
can be rejected. To formalise this iterative procedure, we modify the
filtering count $N(t)$ to exclude features that have
already been rejected. Specifically, for any rejection
set $\cR\subseteq[m]$ and $t\in[0,\alpha]$, define
\begin{equation*}
    N_{\cR}(t)
    :=
    \sum_{i\in[m]\setminus\cR}
    \mathbf{1}\{F_i<t\}.
\end{equation*}
We refer to the sequential extension as AdaFilter-SeqBon
and restate it below in our notation.

\begin{definition}[AdaFilter-SeqBon]
\label{def:sequential_adafilter_bon}
For a target FWER level $\alpha\in(0,1)$,
initialise $\cR_{\mathrm{Bon}}^{(0)}=\varnothing$.
For $k=1,2,\dots$, define the feasible region
\begin{equation*}
    \cT^{(k)}_{\mathrm{SeqBon}} :=
    \left\{
        t\in[0,\alpha]:
        t N_{\cR_{\mathrm{Bon}}^{(k-1)}}(t)
        \leq\alpha
    \right\},
\end{equation*}
and the data-dependent threshold
$\hat{t}^{(k)}_{\mathrm{Bon}}
= \sup \cT^{(k)}_{\mathrm{SeqBon}}$.
Update the rejection set according to
\begin{equation*}
    \cR_{\mathrm{Bon}}^{(k)}
    =
    \cR_{\mathrm{Bon}}^{(k-1)}
    \cup
    \left\{
        i\notin\cR_{\mathrm{Bon}}^{(k-1)}:
        S_i<\hat{t}^{(k)}_{\mathrm{Bon}}
    \right\}.
\end{equation*}
The procedure terminates at the first iteration
$k=k^\star$ for which
$\cR_{\mathrm{Bon}}^{(k)}
=\cR_{\mathrm{Bon}}^{(k-1)}$.
AdaFilter-SeqBon outputs the rejection set
\begin{equation*}
    \cR_{\mathrm{SeqBon}}
    :=
    \cR_{\mathrm{Bon}}^{(k^\star)}.
\end{equation*}
\end{definition}

The first iteration of AdaFilter-SeqBon coincides
with AdaFilter-Bon. At each subsequent iteration,
previously rejected hypotheses are excluded from
the filtering count, so the rejection
threshold cannot decrease. Consequently,
\begin{equation*}
    \cR_{\mathrm{Bon}}
    \subseteq
    \cR_{\mathrm{SeqBon}},
\end{equation*}
showing that AdaFilter-SeqBon uniformly improves
upon AdaFilter-Bon in terms of the rejection set.
Under mutual independence of all base $p$-values,
AdaFilter-SeqBon also controls the FWER at level
$\alpha$ by \citet[Section~S1]{Wang2022}.

\section{Uniformly Improved PC $p$-Values and Filtering Statistics for AdaFilter}
\label{sec:uniform_improved_stat}

Bonferroni PC $p$-values are known to be conservative in comparison to other PC $p$-value constructions \citep{Benjamini2008}. Nevertheless, they have the important advantage of satisfying the conditional validity property in \eqref{conditional_validity}, which underlies the FWER control of AdaFilter-Bon and AdaFilter-SeqBon. The following lemma establishes sufficient conditions under which replacing the Bonferroni PC $p$-values and their filtering statistics yields procedures that uniformly dominate AdaFilter-Bon and AdaFilter-SeqBon in terms of rejections.

\begin{lemma}[Uniform improvement under valid transformations]\label{lem:uniform_improvement_AdaFilter}
Suppose that all base $p$-values $(P_{ij})_{(i,j)\in[m]\times[n]}$ are mutually independent. Let $g:[0,n-u+1]\to[0,1]$ be continuous and strictly increasing, with $g(0)=0$ and $g(x)\leq x$ for all $x\in[0,n-u+1]$. Suppose that each $g(S_i)$ is a valid PC $p$-value and satisfies the following conditional validity
property whenever $H_i^{u/n}$ is true:
\begin{equation}\label{conditional_validity_g}
    \Pr(g(S_i)<t\mid g(F_i)<t)\leq t
\end{equation}
for every $t\in(0,1]$ such that
$\Pr(g(F_i)<t)>0$. Let $\cR_{\mathrm{Bon}}^{(g)}$ and $\cR_{\mathrm{SeqBon}}^{(g)}$ denote the rejection sets produced by applying AdaFilter-Bon and AdaFilter-SeqBon to $\{(g(S_i),g(F_i))\}_{i\in[m]}$, respectively. Then:
\begin{enumerate}
    \item The AdaFilter-Bon procedure based on $\{(g(S_i),g(F_i))\}_{i\in[m]}$ controls the FWER at level $\alpha$ and
    $
        \cR_{\mathrm{Bon}}
        \subseteq
        \cR_{\mathrm{Bon}}^{(g)}.
    $

    \item The AdaFilter-SeqBon procedure based on $\{(g(S_i),g(F_i))\}_{i\in[m]}$ controls the FWER at level $\alpha$ and
    $
        \cR_{\mathrm{SeqBon}}
        \subseteq
        \cR_{\mathrm{SeqBon}}^{(g)}.
    $
\end{enumerate}
\end{lemma}

Consider the following strictly increasing transformation:
\begin{equation*}
    g_{\mathrm{Sid}}(x)
    =
    1-\left(1-\frac{x}{n-u+1}\right)^{n-u+1},
    \qquad x\in[0,n-u+1],
\end{equation*}
where $g_{\mathrm{Sid}}(x)\leq x$ by Bernoulli's inequality and $g_{\mathrm{Sid}}(0) = 0$. For each feature $i$, define the PC $p$-value
\begin{equation*}
    \tilde{S}_i
    :=g_{\mathrm{Sid}}(S_i)
    =
    1-(1-P_{i(u)})^{n-u+1},
\end{equation*}
and the corresponding filtering statistic
\begin{equation*}
    \tilde{F}_i
    :=g_{\mathrm{Sid}}(F_i)
    =
    1-(1-P_{i(u-1)})^{n-u+1}.
\end{equation*}
We refer to $\tilde{S}_i$ and $\tilde{F}_i$ as the \emph{\Sidak\ PC $p$-value} and \emph{\Sidak\ filtering statistic}, respectively, reflecting their connection to the classical \Sidak\ correction; see the proof of \thmref{Sidak_properties}$(i)$ in \appref{Sidak_properties} for details. The following theorem establishes that this construction preserves the validity and conditional validity properties required by AdaFilter.

\begin{theorem}[Validity and conditional validity of \Sidak\ PC $p$-values]
\label{thm:Sidak_properties}
Suppose that the base $p$-values $P_{i1},\dots,P_{in}$ are mutually independent. Then, whenever $H_i^{u/n}$ is true:
\begin{enumerate}
    \item $\tilde{S}_i$ is a valid PC $p$-value, that is,
    \begin{equation*}
        \Pr(\tilde{S}_i\leq t)\leq t
        \qquad\text{for all $t\in[0,1]$}.
    \end{equation*}

    \item $\tilde{S}_i$ is conditionally valid given the filtering event determined by $\tilde{F}_i$, that is,
    \begin{equation*}
        \Pr(\tilde{S}_i<t\mid\tilde{F}_i<t)
        \leq t
    \end{equation*}
    for every $t\in(0,1]$ such that
    $\Pr(\tilde{F}_i<t)>0$.
\end{enumerate}
\end{theorem}

Combining \thmref{Sidak_properties} with \lemref{uniform_improvement_AdaFilter} immediately yields the following result.

\begin{corollary}[Uniform improvement of AdaFilter using \Sidak\ statistics]
\label{cor:uniform_improvement}
Suppose that all base $p$-values $(P_{ij})_{(i,j)\in[m]\times[n]}$ are mutually independent. Let $\tilde{\cR}_{\mathrm{Bon}}$ and $\tilde{\cR}_{\mathrm{SeqBon}}$ denote the rejection sets outputted by applying AdaFilter-Bon and AdaFilter-SeqBon to
$\{(\tilde{S}_i,\tilde{F}_i)\}_{i\in[m]}$, respectively. Then both resulting procedures control the FWER at level $\alpha$, and
\begin{equation*}
    \cR_{\mathrm{Bon}}
    \subseteq
    \tilde{\cR}_{\mathrm{Bon}}
    \qquad\text{and}\qquad
    \cR_{\mathrm{SeqBon}}
    \subseteq
    \tilde{\cR}_{\mathrm{SeqBon}}.
\end{equation*}
Hence, the \Sidak-based AdaFilter-Bon and AdaFilter-SeqBon procedures uniformly dominate their respective Bonferroni counterparts in terms of rejection sets.
\end{corollary}

It is natural to ask whether the \Sidak\ transformation
can be improved further to yield uniformly smaller
PC $p$-values while preserving the validity and
conditional validity properties required by AdaFilter.
Let $\cG$ denote the class of continuous, strictly
increasing functions $g:[0,n-u+1]\to[0,1]$ with
$g(0)=0$ and $g(x)\leq x$ for all
$x\in[0,n-u+1]$, such that, for every collection
of mutually independent valid base $p$-values,
$g(S_i)$ is a valid PC $p$-value and satisfies
the conditional validity property in
\eqref{conditional_validity_g} whenever
$H_i^{u/n}$ is true.

The following theorem provides a complete characterisation
of $\cG$ and establishes that the \Sidak\ transformation
is pointwise optimal among transformations applied jointly
to the Bonferroni PC $p$-values and their corresponding
filtering statistics that satisfy the conditions defining
$\cG$.

\begin{theorem}[Pointwise optimality of the \Sidak\ transformation]
\label{thm:sidak_g_optimal}
Let $g:[0,n-u+1]\to[0,1]$ be continuous and
strictly increasing, with $g(0)=0$ and
$g(x)\leq x$ for all $x\in[0,n-u+1]$.
Then $g\in\cG$ if and only if
\begin{equation*}
    g_{\mathrm{Sid}}(x)\leq g(x)
    \qquad\text{for all $x\in[0,n-u+1]$}.
\end{equation*}
Consequently, $g_{\mathrm{Sid}}\in\cG$ and is the
pointwise smallest transformation in $\cG$.
\end{theorem}

The pointwise optimality of $g_{\mathrm{Sid}}$ also
translates into optimality in terms of the rejection
sets produced by AdaFilter-Bon and AdaFilter-SeqBon,
as established in the following corollary.

\begin{corollary}[Rejection-set optimality of the \Sidak\ transformation]
\label{cor:sidak_rejection_optimal}
For every $g\in\cG$,
\begin{equation*}
    \cR_{\mathrm{Bon}}^{(g)}
    \subseteq
    \cR_{\mathrm{Bon}}^{(g_{\mathrm{Sid}})}
    \qquad\text{and}\qquad
    \cR_{\mathrm{SeqBon}}^{(g)}
    \subseteq
    \cR_{\mathrm{SeqBon}}^{(g_{\mathrm{Sid}})}.
\end{equation*}
Hence, the \Sidak\ transformation uniformly
maximises the rejection sets of AdaFilter-Bon
and AdaFilter-SeqBon among all transformations
in $\cG$.
\end{corollary}

\section{An Adaptive \Sidak\ Correction for FWER Control Under the AdaFilter Framework}
\label{sec:sidak_fwer}

Under independence, the classical \Sidak\ correction
uniformly improves upon the Bonferroni correction
for ordinary FWER control. In particular, when testing
$m$ hypotheses, the \Sidak\ rejection threshold
\begin{equation*}
    1-(1-\alpha)^{1/m}
\end{equation*}
is at least as large as the Bonferroni rejection
threshold $\alpha/m$.
This motivates replacing the post-filter Bonferroni
multiplicity correction within the AdaFilter framework
by its \Sidak\ analogue, with the aim of further
enlarging the rejection set.

For any rejection set $\cR\subseteq[m]$ and $t\in[0,\alpha]$, define
\begin{equation*}
    \tilde{N}_{\cR}(t)
    :=
    \sum_{i\in[m]\setminus\cR}
    \mathbf{1}\{\tilde{F}_i<t\}
    \quad \text{and} \quad
    \tilde{N}(t)
    :=
    \tilde{N}_{\varnothing}(t)
    =
    \sum_{i\in[m]}
    \mathbf{1}\{\tilde{F}_i<t\}.
\end{equation*}
Thus, $\tilde{N}_{\cR}(t)$ counts the number of unrejected features that pass the filtering step based on $(\tilde{F}_i)_{i\in[m]}$ at threshold $t$. We use this quantity to define the \Sidak\ analogue of AdaFilter-Bon.

\begin{definition}[AdaFilter-Sid]
\label{def:adafilter_sidak}
For a target FWER level $\alpha\in(0,1)$, define the feasible region
\begin{equation*}
    \cT_{\mathrm{Sid}}
    :=
    \left\{
        t\in[0,\alpha]:
        1-(1-t)^{\tilde{N}(t)}
        \leq\alpha
    \right\}
\end{equation*}
and the data-dependent threshold
$\hat{t}_{\mathrm{Sid}}:=\sup\cT_{\mathrm{Sid}}$.
AdaFilter-Sid outputs the rejection set
\begin{equation*}
    \cR_{\mathrm{Sid}}
    :=
    \left\{
        i\in[m]:
        \tilde{S}_i<\hat{t}_{\mathrm{Sid}}
    \right\}.
\end{equation*}
\end{definition}

As in the Bonferroni case, further gains can be obtained by updating the multiplicity correction after each round of rejections. This leads to a sequential version of AdaFilter-Sid.

\begin{definition}[AdaFilter-SeqSid]
\label{def:sequential_adafilter_sidak}
For a target FWER level $\alpha\in(0,1)$,
initialise $\cR_{\mathrm{Sid}}^{(0)}=\varnothing$.
For $k=1,2,\dots$, define the feasible region
\begin{equation*}
    \cT_{\mathrm{SeqSid}}^{(k)}
    :=
    \left\{
        t\in[0,\alpha]:
        1-(1-t)^{
            \tilde{N}_{\cR_{\mathrm{Sid}}^{(k-1)}}(t)
        }
        \leq\alpha
    \right\},
\end{equation*}
and the data-dependent threshold
$\hat{t}_{\mathrm{Sid}}^{(k)}
:= \sup\cT_{\mathrm{SeqSid}}^{(k)}$.
Update the rejection set according to
\begin{equation*}
    \cR_{\mathrm{Sid}}^{(k)}
    =
    \cR_{\mathrm{Sid}}^{(k-1)}
    \cup
    \left\{
        i\notin\cR_{\mathrm{Sid}}^{(k-1)}:
        \tilde{S}_i<\hat{t}_{\mathrm{Sid}}^{(k)}
    \right\}.
\end{equation*}
The procedure terminates at the first iteration
$k=k^\star$ for which
$\cR_{\mathrm{Sid}}^{(k)}
=\cR_{\mathrm{Sid}}^{(k-1)}$.
AdaFilter-SeqSid outputs the rejection set
\begin{equation*}
    \cR_{\mathrm{SeqSid}}
    :=
    \cR_{\mathrm{Sid}}^{(k^\star)}.
\end{equation*}
\end{definition}

The following theorem establishes that the proposed procedures retain FWER control despite the data-dependent filtering step, while uniformly improving upon their Bonferroni-based counterparts.

\begin{theorem}[FWER control and uniform dominance of \Sidak-based adaptive filtering procedures]\label{thm:sidak_fwer}
Suppose that all base $p$-values
$\{P_{ij}\}_{(i,j)\in[m]\times[n]}$
are mutually independent and valid. Then, for any $\alpha\in(0,1)$:
\begin{enumerate}
    \item AdaFilter-Sid controls the FWER at level $\alpha$ and
    $
        \tilde{\cR}_{\mathrm{Bon}}
        \subseteq
        \cR_{\mathrm{Sid}}.
    $

    \item AdaFilter-SeqSid controls the FWER at level $\alpha$
    and
    $
        \tilde{\cR}_{\mathrm{SeqBon}}
        \subseteq
        \cR_{\mathrm{SeqSid}}.
    $
\end{enumerate}
\end{theorem}

Combining \corref{uniform_improvement} and \thmref{sidak_fwer}, and noting that the sequential procedure contains all rejections from its first iteration, we obtain
\begin{equation*}
    \cR_{\mathrm{Bon}}
    \subseteq
    \tilde{\cR}_{\mathrm{Bon}}
    \subseteq
    \cR_{\mathrm{Sid}}
    \subseteq
    \cR_{\mathrm{SeqSid}},
\end{equation*}
and
\begin{equation*}
    \cR_{\mathrm{SeqBon}}
    \subseteq
    \tilde{\cR}_{\mathrm{SeqBon}}
    \subseteq
    \cR_{\mathrm{SeqSid}}.
\end{equation*}
These inclusions highlight two distinct sources of improvement:
replacing the Bonferroni PC $p$-values and filtering
statistics by their \Sidak\ counterparts, and replacing
the post-filter Bonferroni multiplicity correction
by the \Sidak\ correction.
Each improvement can only enlarge the rejection set,
and their combination yields procedures that uniformly
dominate their Bonferroni counterparts, both with and
without sequential rejection.

\section{Simulation studies}
\label{sec:simulations}

We conduct simulation studies to evaluate the FWER control
and power of the proposed procedures in comparison with
existing methods for testing PC null hypotheses. Throughout,
the nominal FWER level is $\alpha=0.05$ for simultaneously
testing $H_1^{u/n},\dots,H_m^{u/n}$, where the underlying
base null hypotheses are
\begin{equation*}
    H_{ij}: \theta_{ij}=0,
    \qquad (i,j)\in[m]\times[n].
\end{equation*}
Let $Z_{ij}$ denote the test statistic corresponding
to $H_{ij}$, which follows the standard normal
distribution whenever $H_{ij}$ is true.
The associated two-sided base $p$-value is
computed as
\begin{equation*}
    P_{ij}=2\Phi(-|Z_{ij}|),
\end{equation*}
where $\Phi$ denotes the standard normal cumulative
distribution function.
The distributions of the test statistics under
the alternatives are specified separately for
each simulation setting.

We consider the following eight FWER-controlling
procedures, the first six of which are variants of AdaFilter:
\begin{itemize}
    \item \textbf{AdaFilter-Bon}: The procedure in
    \defref{adafilter_bon};

    \item \textbf{AdaFilter-SeqBon}: The sequential procedure
    in \defref{sequential_adafilter_bon};

    \item \textbf{AdaFilter-Bon (\Sidak\ PC)}: The procedure
    in \defref{adafilter_bon} but applied to
    $\{(\tilde{S}_i,\tilde{F}_i)\}_{i\in[m]}$;

    \item \textbf{AdaFilter-SeqBon (\Sidak\ PC)}:
    The procedure in \defref{sequential_adafilter_bon}
    but applied to
    $\{(\tilde{S}_i,\tilde{F}_i)\}_{i\in[m]}$;

    \item \textbf{AdaFilter-Sid}: The procedure in
    \defref{adafilter_sidak};

    \item \textbf{AdaFilter-SeqSid}: The sequential procedure
    in \defref{sequential_adafilter_sidak};

    \item \textbf{CoFilter-Holm}: The CoFilter procedure
    of \citet{Dickhaus2026}, which uses Fisher PC
    $p$-values for filtering at the prespecified threshold
    $\tau=0.1$, as considered by \citet{Dickhaus2026},
    and applies Holm's procedure to the resulting
    filter-adjusted (conditional) PC $p$-values.
    Specifically, the Fisher PC $p$-value for testing
    $H_i^{u/n}$ is given by
    \begin{equation*}
        S_i^{\operatorname{Fisher}}
        =
        1-F_{\chi^2_{2(n-u+1)}}\left(
            -2\sum_{j=u}^n\log P_{i(j)}
        \right),
    \end{equation*}
    where $F_{\chi^2_{2(n-u+1)}}$ denotes the cumulative
    distribution function of a chi-squared random
    variable with $2(n-u+1)$ degrees of freedom;

    \item \textbf{Hommel-Fisher}: Hommel's FWER-controlling
    procedure \citep{Hommel1988} applied directly to the Fisher PC
    $p$-values $\{S_i^{\operatorname{Fisher}}\}_{i\in[m]}$.
\end{itemize}

All eight procedures provide FWER control under
mutual independence of the base $p$-values
$(P_{ij})_{(i,j)\in[m]\times[n]}$, with CoFilter-Holm
additionally requiring that the $p$-values
corresponding to true base null hypotheses are
uniformly distributed.
The first seven procedures exploit filtering,
whereas Hommel-Fisher serves as a non-filtering
benchmark.

\subsection{Simulation setting 1}
\label{sec:sim_mixture}

We first investigate the empirical performance of the
eight competing procedures under a simple two-component
normal mixture model. We consider $m=500$ features
and the following six configurations of the replicability
level $u$ and the number of studies $n$:
\begin{equation*}
    (u,n)\in
    \{(2,2),(2,3),(3,3),(2,4),(3,4),(4,4)\}.
\end{equation*}
For each $(i,j)\in[m]\times[n]$, independently generate
\begin{equation*}
    \theta_{ij}\sim
    0.5\delta_0+0.5\delta_\mu,
    \qquad
    Z_{ij}\mid\theta_{ij}
    \sim\mathrm{N}(\theta_{ij},1),
\end{equation*}
where $\delta_0$ denotes the point mass at $0$ and
$\mu\in\{2,3,4,5\}$ controls the signal strength.
The PC null hypothesis $H_i^{u/n}$ is false if at
least $u$ of the underlying effects
$\theta_{i1},\dots,\theta_{in}$ are nonzero.

\begin{figure}
    \centering
    \includegraphics[width=0.9\linewidth]
    {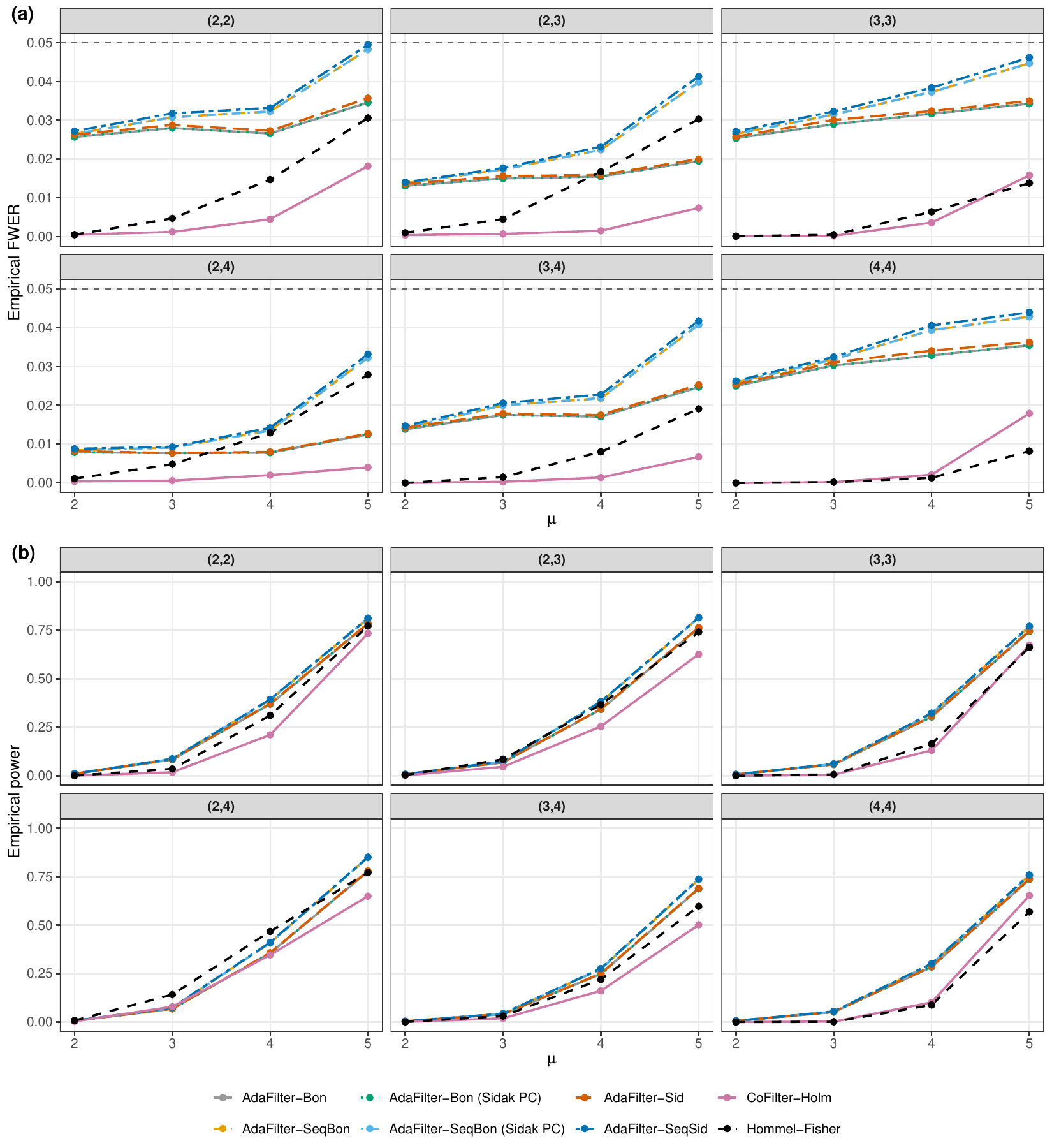}
    \caption{Empirical FWER and power of the eight
    competing procedures under the independent
    two-point normal mixture model, based on
    $10,000$ simulation repetitions.
    Panel (a) reports the empirical FWER and
    panel (b) reports the empirical power.
    Each panel contains the six $(u,n)$ configurations,
    with signal strength $\mu\in\{2,3,4,5\}$.
    In panel (a), the horizontal dashed line
    indicates the nominal FWER level $\alpha=0.05$.}
    \label{fig:additional_fwer_power}
\end{figure}

For each simulation setting, we perform $10,000$
independent repetitions. Empirical FWER is estimated
as the proportion of repetitions in which at least
one true PC null hypothesis is rejected, whereas
empirical power is estimated by averaging the
proportion of false PC null hypotheses rejected
over the repetitions. The empirical FWER and power results are
presented in Figure~\ref{fig:additional_fwer_power}(a)
and (b), respectively.

\begin{figure}
    \centering
    \includegraphics[width=0.9\linewidth]
    {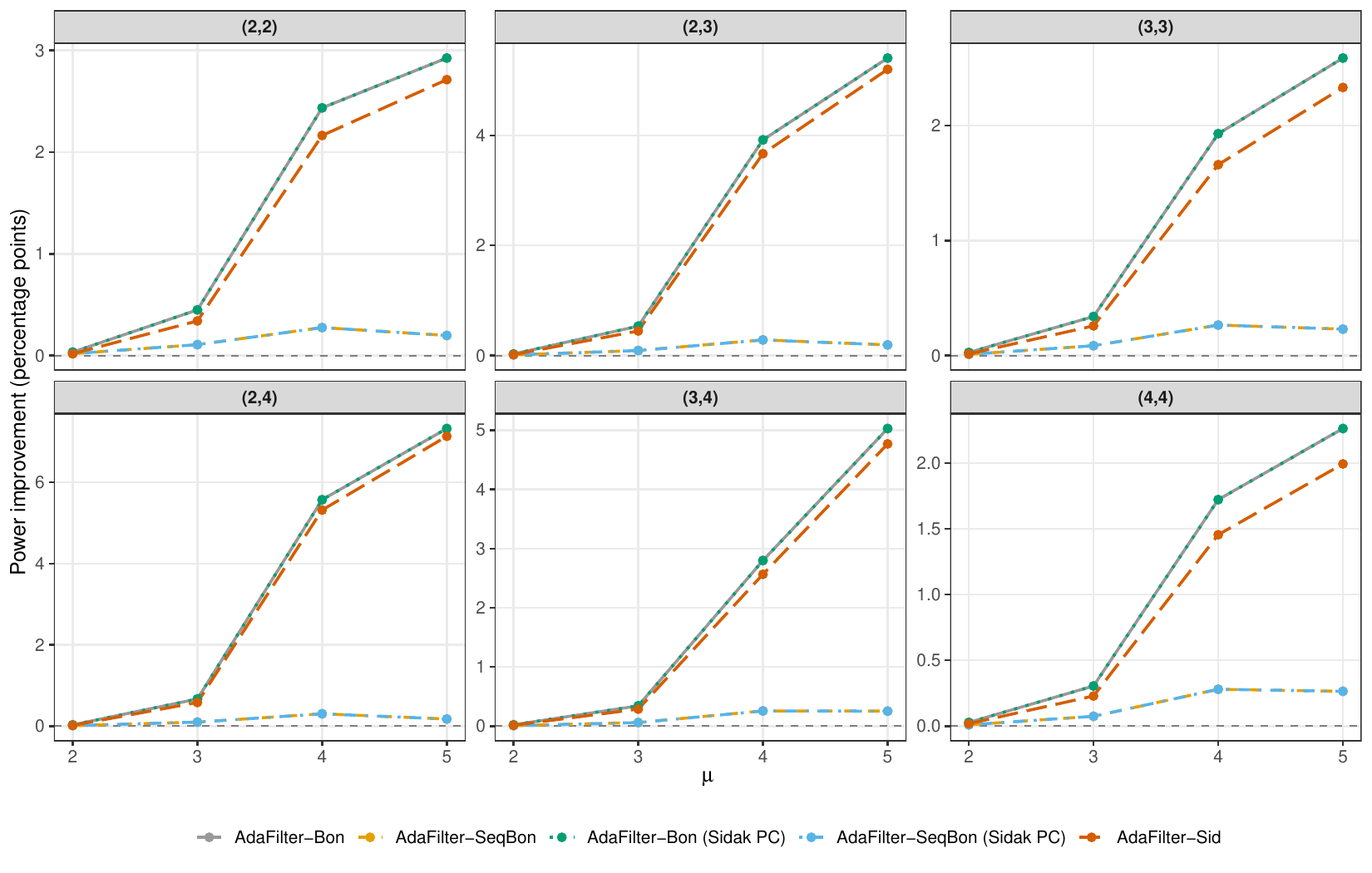}
    \caption{Empirical power improvements (in
    percentage points) of AdaFilter-SeqSid over
    the other five AdaFilter variants under the
    independent two-point normal mixture model,
    based on $10,000$ simulation repetitions.
    Each panel corresponds to a different $(u,n)$
    configuration, with signal strength
    $\mu\in\{2,3,4,5\}$.
    The vertical axes have different scales across
    panels to facilitate comparison of the power
    improvements.}
    \label{fig:additional_power_gain}
\end{figure}

To examine the power improvements of the proposed
procedures in this paper more closely, we use AdaFilter-SeqSid as
a benchmark. As established by our theoretical
results in \secref{sidak_fwer}, AdaFilter-SeqSid uniformly dominates all
other AdaFilter variants in terms of their rejection
sets, making it a natural reference for comparison.
For each remaining AdaFilter variant
\begin{equation*}
    \mathcal{M}\in
    \left\{
    \begin{array}{l}
        \text{AdaFilter-Bon},\\
        \text{AdaFilter-SeqBon},\\
        \text{AdaFilter-Bon (\Sidak\ PC)},\\
        \text{AdaFilter-SeqBon (\Sidak\ PC)},\\
        \text{AdaFilter-Sid}
    \end{array}
    \right\},
\end{equation*}
we define the AdaFilter-SeqSid empirical power \textit{improvement}
(in percentage points) as
\begin{equation*}
    \Delta_{\mathcal{M}}
    =
    100\left(
        \widehat{\mathrm{Power}}_{\mathrm{SeqSid}}
        -
        \widehat{\mathrm{Power}}_{\mathcal{M}}
    \right),
\end{equation*}
where $\widehat{\mathrm{Power}}_{\mathrm{SeqSid}}$
and $\widehat{\mathrm{Power}}_{\mathcal{M}}$
denote the empirical powers of AdaFilter-SeqSid
and $\mathcal{M}$, respectively.
The results are presented in
\figref{additional_power_gain}.

The following observations from the simulation are made:
\begin{itemize}
    \item As shown in Figure~\ref{fig:additional_fwer_power}(b),
    empirical power generally increases with signal
    strength $\mu$ across all procedures. Among the six
    AdaFilter variants, AdaFilter-SeqSid consistently
    achieves the highest power, in agreement with its
    theoretical uniform dominance. The sequential AdaFilter variants also generally
outperform their non-sequential counterparts,
demonstrating the benefit of sequential rejection.

    \item The power curves of the six AdaFilter variants
    in Figure~\ref{fig:additional_fwer_power}(b) are often
    very close, making their differences difficult to
    discern visually. These differences become clearer
    in \figref{additional_power_gain}, which directly
    displays the power improvements of AdaFilter-SeqSid
    over the other five variants. The improvements are
    consistently nonnegative and generally become more
    pronounced as $\mu$ increases, particularly relative
    to the non-sequential variants. 

    \item Figure~\ref{fig:additional_fwer_power}(a)
    shows that all eight procedures maintain empirical
    FWER control at the nominal level $\alpha=0.05$
    across all simulation settings. In particular,
    the proposed \Sidak-based AdaFilter procedures
    maintain FWER control despite their less
    conservative rejection thresholds.
\end{itemize}

Overall, these results provide empirical support for
the proposed improvements under a conventional
simulation model. Although the power gains are
generally modest, the theoretical uniform dominance
guarantees that they are nonnegative without
compromising finite-sample FWER control.
Additional simulation results under positive and
negative AR(1) dependence are presented in
\appref{additional_simulations_dep} to assess the
robustness of the proposed procedures when the
mutual independence assumption is violated.

\subsection{Simulation setting 2}
\label{sec:sim_calibrated}

The preceding simulation results demonstrate that the
proposed improvements generally yield modest power
gains under a conventional normal mixture model.
We now construct a deliberately calibrated setting
to illustrate that even a small improvement in the
rejection threshold can translate into a substantial
power improvement when PC $p$-values are highly
concentrated near the decision boundary.

We consider a replicability analysis involving
$m=100$ features across $n=5$ independent studies.
We focus on the replicability level $u=3$ and
test the PC null hypotheses
$H^{3/5}_1,\dots,H^{3/5}_{100}$.

For each $(i,j)\in[100]\times[5]$,
we independently generate
\begin{equation*}
    Z_{ij}\sim\mathrm{N}(\theta_{ij},\sigma_{ij}^2),
    \qquad
    (\theta_{ij},\sigma_{ij}^2)
    =
    \begin{cases}
        (6,0.01),
        & \text{if $i\in[10]$ and $j\in\{1,2\}$},\\
        (c,v),
        & \text{if $i\in[10]$ and $j=3$},\\
        (0,1),
        & \text{otherwise},
    \end{cases}
\end{equation*}
where $c$ and $v$ specify the mean and variance
of the test statistics for the first $10$ features
in the third study, respectively.
We fix $c=3.14010315573$ and consider
\begin{equation*}
    v\in\{10^{-5},1,5,10,15\}.
\end{equation*}
By construction, the first $10$ features have
exactly three non-null base hypotheses, whereas
all base hypotheses are null for the remaining
$90$ features. Consequently, the first $10$
PC null hypotheses are false and the remaining
$90$ are true, i.e.,
$|\cH_0^{3/5}|=90$.

To motivate the choice of $c$, consider
AdaFilter-Bon (\Sidak\ PC) and AdaFilter-Sid,
which use the same \Sidak\ PC $p$-values and
filtering statistics but differ in their
post-filter multiplicity corrections.
For $u=3$ and $n=5$, these statistics are
\begin{equation*}
    \tilde{S}_i
    =
    1-(1-P_{i(3)})^3
    \quad\text{and}\quad
    \tilde{F}_i
    =
    1-(1-P_{i(2)})^3.
\end{equation*}

For the first $10$ features, the signals
in the first two studies are sufficiently
strong that their filtering statistics
are approximately zero. In contrast,
the remaining $90$ features rarely pass
the filtering step at the relevant thresholds.
Consequently, approximately $10$ features
pass the filtering step, yielding the
approximate rejection thresholds
\begin{equation}\label{approx_thresholds}
    \tilde{t}_{\mathrm{Bon}}
    \approx
    \frac{\alpha}{10}
    =0.005
    \quad\text{and}\quad
    \hat{t}_{\mathrm{Sid}}
    \approx
    1-(1-\alpha)^{1/10}
    \approx0.005116
\end{equation}
for AdaFilter-Bon (\Sidak\ PC) and
AdaFilter-Sid, respectively.

Under this approximation, the two procedures
can produce different rejection decisions
when a \Sidak\ PC $p$-value lies between
their rejection thresholds.
To construct a setting in which this occurs
with high probability, we consider the
midpoint of the approximate thresholds
in \eqref{approx_thresholds}:
\begin{equation*}
    t^\star
    =
    \frac{1}{2}
    \left(
        \frac{\alpha}{10}
        +
        1-(1-\alpha)^{1/10}
    \right)
    \approx0.005058.
\end{equation*}

For the first $10$ features, the base
$p$-values from the first two studies
are approximately zero, giving
\begin{equation*}
    P_{i(3)}
    \approx
    \min\{P_{i3},P_{i4},P_{i5}\}.
\end{equation*}
When $v$ is sufficiently small,
$P_{i3}$ is concentrated near
$2\Phi(-c)$. If this value is also
sufficiently small, the null-study
$p$-values $P_{i4}$ and $P_{i5}$
are unlikely to be smaller than
$P_{i3}$. Thus,
$P_{i(3)}\approx2\Phi(-c)$
with high probability, giving
\begin{equation*}
    \tilde{S}_i
    \approx
    1-\left(1-2\Phi(-c)\right)^3.
\end{equation*}

We therefore calibrate $c$ by solving
\begin{equation*}
    1-\left(1-2\Phi(-c)\right)^3
    =
    t^\star,
\end{equation*}
which yields
\begin{equation*}
    c
    =
    \Phi^{-1}\left(
        1-\frac{1-(1-t^\star)^{1/3}}{2}
    \right)
    \approx3.140103.
\end{equation*}

When $v=10^{-5}$, this choice concentrates
the \Sidak\ PC $p$-values for the first
$10$ features near $t^\star$, which lies
strictly between the two approximate
rejection thresholds in \eqref{approx_thresholds}.
This creates a setting in which AdaFilter-Sid
can achieve substantially higher power than
AdaFilter-Bon (\Sidak\ PC), even though their
rejection thresholds differ only slightly.
The remaining values of $v$ allow us to
investigate how this improvement changes
as the variability of the test statistics
increases.

\begin{figure}[t]
    \centering
    \includegraphics[width=0.90\linewidth]{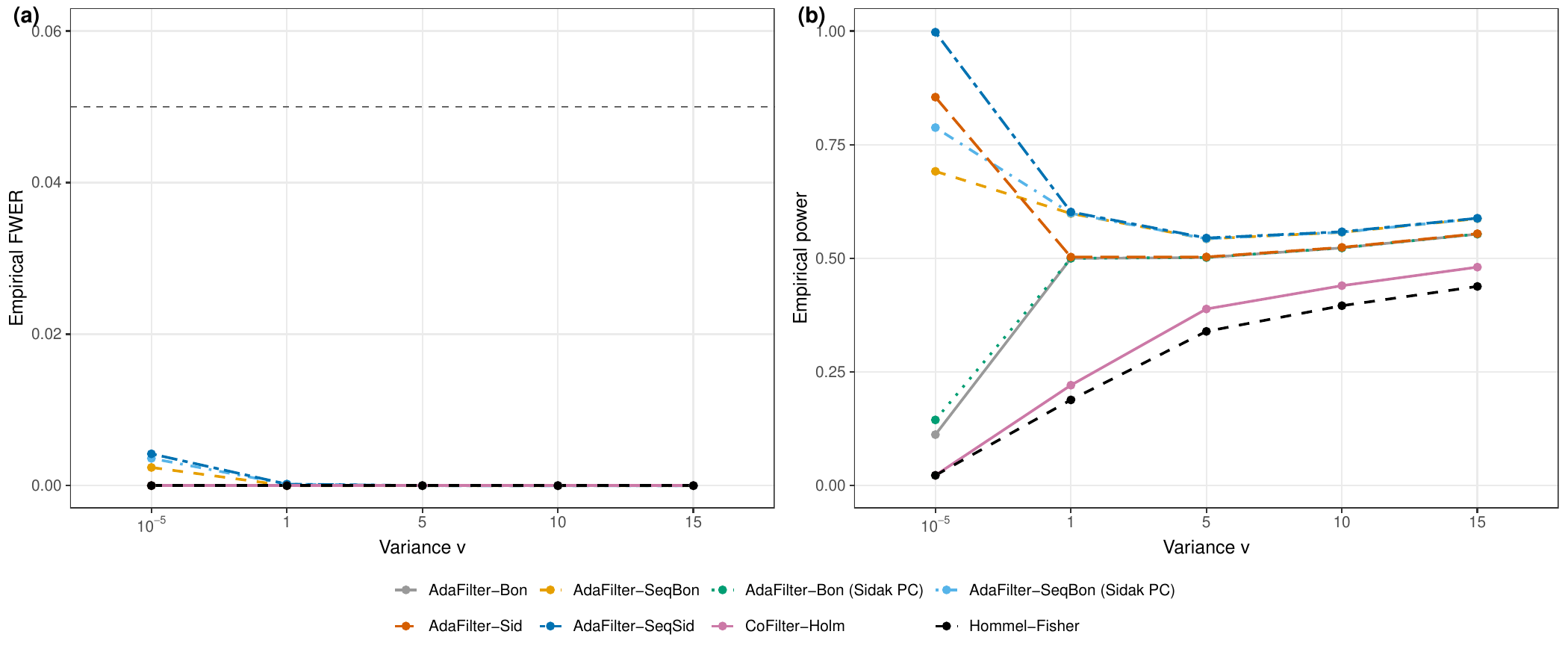}
    \caption{Empirical FWER (panel (a)) and power (panel (b))
    of the eight competing procedures as functions of the
    variance parameter $v$ under the calibrated setting
    $c=3.1401031557$. The replicability level is $u=3$
    out of $n=5$ studies, with $m=100$ features and
    nominal FWER level $\alpha=0.05$.
    The horizontal dashed line in panel (a) indicates
    the nominal FWER level. All results are based on
    $10{,}000$ Monte Carlo repetitions.}
    \label{fig:sim_results}
\end{figure}

We apply the eight competing procedures at the nominal
FWER level $\alpha=0.05$. As in Simulation setting 1,
empirical FWER and power are estimated from $10,000$
Monte Carlo repetitions. The results are presented in \figref{sim_results}.

The following observations are made:
\begin{itemize}
    \item \textbf{Substantial power improvements.}
    When $v=10^{-5}$, the power differences among the
    AdaFilter variants become particularly pronounced.
    As explained above, this setting concentrates the
    \Sidak\ PC $p$-values for the first $10$ features
    near the rejection thresholds of AdaFilter-Bon
    (\Sidak\ PC) and AdaFilter-Sid, allowing the
    sharper \Sidak\ multiplicity correction to yield
    substantial power gains. In this setting,
    AdaFilter-SeqSid achieves power close to one,
    compared with approximately $0.85$ for
    AdaFilter-Sid, $0.80$ for AdaFilter-SeqBon
    (\Sidak\ PC), $0.70$ for AdaFilter-SeqBon,
    and only $0.12$ for AdaFilter-Bon.
    In particular, the substantial improvement of
    AdaFilter-SeqSid over AdaFilter-SeqBon
    (\Sidak\ PC) illustrates how the sharper
    \Sidak\ multiplicity correction can translate
    into considerable power gains when PC $p$-values
    are concentrated near the rejection thresholds.
    Its additional improvement over AdaFilter-Sid
    illustrates the benefit of sequential rejection.

    \item \textbf{Sensitivity to variability.}
    As $v$ increases, the \Sidak\ PC $p$-values
    become less concentrated near the rejection
    thresholds, and the power differences among
    the AdaFilter variants diminish.
    For example, when $v=15$, AdaFilter-SeqSid
    achieves power $0.5877$, compared with $0.5865$
    for AdaFilter-SeqBon and $0.5867$ for
    AdaFilter-SeqBon (\Sidak\ PC).
    Thus, although the power gains are considerably
    smaller in this setting, AdaFilter-SeqSid
    remains the most powerful among the AdaFilter
    variants, consistent with the uniform dominance
    results established in \corref{uniform_improvement}
    and \thmref{sidak_fwer}.
    In comparison, CoFilter-Holm and Hommel-Fisher
    achieve lower powers of $0.4790$ and $0.4370$,
    respectively.

    \item \textbf{FWER control.}
    All eight procedures maintain empirical FWER
    below the nominal level $\alpha=0.05$ across
    the considered values of $v$, consistent with
    their theoretical FWER guarantees.
\end{itemize}

Taken together, the two simulation settings
demonstrate that the proposed procedures yield
systematic power improvements under a conventional
normal mixture model, while the calibrated setting
shows that even a small increase in the rejection
threshold can translate into a substantial power
gain when PC $p$-values are highly concentrated
near the decision boundary.

\section{Real Data}\label{sec:real_data}

We illustrate the proposed procedures using the publicly
available microarray dataset GSE30495 from the Gene
Expression Omnibus (GEO) \citep{Georgiadi2012}.
The dataset contains gene expression measurements from
the hearts of wild-type and PPAR$\alpha$ knockout mice
subjected to four different fatty acid treatments
(oleic acid, linoleic acid, linolenic acid, and
docosahexaenoic acid) and a synthetic PPAR$\alpha$
agonist (Wy14,643).
We therefore consider five treatment-specific comparisons
between wild-type and knockout mice.
For consistency with our methodological framework,
we refer to each comparison as a study, although all
five comparisons originate from the same experiment.
Since the comparisons involve disjoint sets of mice,
we treat the five studies as independent.
The analysis comprises $m=45{,}101$ gene-expression
features (microarray probe sets) across $n=5$ studies.
For each feature, the objective is to identify
differential expression replicated in at least $u$
of the five studies, where $u\in\{2,3,4,5\}$.

To construct the base $p$-values, raw microarray data
were independently preprocessed within each study
using robust multi-array average (RMA) normalisation
\citep{Irizarry2003}.
Two-sided $p$-values for differential expression
were then obtained from ordinary $t$-statistics
computed using study-specific linear models fitted
with the \texttt{limma} package
\citep{Smyth2004,Ritchie2015}.
The analyses were performed separately using disjoint
samples, without sharing variance estimates across
studies.
We applied the six AdaFilter procedures considered
in \secref{simulations} at the nominal FWER level
$\alpha=0.05$.

\tabref{GSE30495} reports the numbers of rejected
PC null hypotheses.
The sequential procedures consistently yield
substantially more rejections than their
non-sequential counterparts.
Among all methods, AdaFilter-SeqSid achieves
the largest number of rejections for every
replicability level (allowing ties).
In particular, compared with AdaFilter-SeqBon,
it identifies three additional replicated signals
for $u=3$ and one additional signal for $u=5$.
These results illustrate the practical benefits
of sequential rejection and the additional, albeit
modest, improvements obtained through the
\Sidak\ correction.

We acknowledge that gene expression measurements
may exhibit dependence across features within
each study, potentially violating the mutual
independence assumption underlying our theoretical
FWER guarantees.
Nevertheless, the additional simulations under
positive and negative AR(1) dependence in
\appref{additional_simulations_dep} provide empirical
evidence that the AdaFilter variants maintain
FWER control under the dependence structures
considered.

\begin{table}[t]
    \centering
    \caption{Numbers of rejected PC null hypotheses for different replicability levels in the GSE30495 dataset ($m=45{,}101$, $n=5$) at $\alpha=0.05$. The largest number of rejections for each replicability level is highlighted in bold.}
    \label{tab:GSE30495}
    \begin{tabular}{lcccc}
        \toprule
        Method & $2/5$ & $3/5$ & $4/5$ & $5/5$ \\
        \midrule
        AdaFilter-Bon                 & 164 & 144 & 129 & 104 \\
        AdaFilter-SeqBon              & \textbf{185} & 175 & \textbf{157} & 138 \\
        AdaFilter-Bon (\Sidak\ PC)    & 164 & 144 & 129 & 104 \\
        AdaFilter-SeqBon (\Sidak\ PC) & \textbf{185} & 175 & \textbf{157} & 138 \\
        AdaFilter-Sid               & 165 & 145 & 129 & 105 \\
        AdaFilter-SeqSid            & \textbf{185} & \textbf{178} & \textbf{157} & \textbf{139} \\
        \bottomrule
    \end{tabular}
\end{table}

\section{Discussion}\label{sec:discussion}

In this paper, we developed \Sidak-based improvements
to the AdaFilter framework for PC testing with FWER
control. A key theoretical contribution is the
establishment of conditional validity for the
\Sidak\ PC $p$-values, allowing them to replace
their Bonferroni counterparts without compromising
FWER control after filtering.
We further established the pointwise optimality of
the \Sidak\ transformation within a class of valid
transformations and developed adaptive \Sidak\
multiplicity corrections that maintain finite-sample
FWER control. The simulation results suggest that
the resulting power improvements are generally modest
but can be substantial when PC $p$-values are highly
concentrated near the rejection threshold.

Several directions for future research are worth exploring.
One is to investigate whether the proposed \Sidak\
improvements can be incorporated into AdaFilter
procedures targeting other error rates, such as the
false discovery rate (FDR) and false discovery
exceedance (FDX) probability, or extended to
directional replicability inference.
Another interesting direction is to extend the
AdaFilter framework, together with the proposed
\Sidak\ improvements, to simultaneous false
discovery proportion (FDP) control
\citep{Goeman2011,Blanchard2020,Katsevich2020}.
A key theoretical challenge would be to exploit
the conditional validity property underlying
AdaFilter to construct simultaneous FDP bounds
that remain valid after adaptive filtering.

\appendix
\section{Technical Lemmas}
In this appendix, we provide and prove technical lemmas that we will leverage to prove the theoretical results presented in the main manuscript.

\begin{lemma}[Properties of the \Sidak\  combining function]\label{lem:sidak_combine}
For any positive integer $k$, define the \Sidak\  combining function
\begin{equation*}
    f_{\mathrm{Sid},k}(P_1,\dots,P_k)
    :=
    1-(1-P_{(1)})^k, \quad (P_1,\dots,P_k) \in [0,1]^k,
\end{equation*}
where $P_{(1)}:=\min_{j\in[k]} P_j$.  Then 
\begin{enumerate}
    \item $f_{\mathrm{Sid},k}$ is symmetric in its arguments;
    \item $f_{\mathrm{Sid},k}$ is nondecreasing in each of its arguments;
    \item if $P_1,\dots,P_k$ are mutually independent valid $p$-values for testing generic null hypotheses $H_1,\dots,H_k$, then $f_{\mathrm{Sid},k}(P_1,\dots,P_k)$ is a valid $p$-value for testing the intersection null hypothesis
    \begin{equation*}
        H^{\star} = \bigcap_{j\in[k]} H_j.
    \end{equation*}
\end{enumerate}
\end{lemma}

\begin{proof}[Proof of \lemref{sidak_combine}$(i)$]
Let $(\pi_1,\dots,\pi_k)$ be any permutation of $(1,\dots,k)$. Then, for any $(P_1,\dots,P_k) \in [0,1]^k$, we have that
\begin{equation*}
     f_{\mathrm{Sid},k}(P_{\pi_1},\dots,P_{\pi_k})  = 1-(1-P_{(1)})^k = f_{\mathrm{Sid},k}(P_{1},\dots,P_{k}).
\end{equation*}
Hence, $f_{\mathrm{Sid},k}$ is symmetric in its arguments.
\end{proof}

\begin{proof}[Proof of \lemref{sidak_combine}$(ii)$]
Consider any $(P'_1,\dots,P'_k),(P_1,\dots,P_k) \in [0,1]^k$ such that $P'_j \geq P_j$ for all $j \in [k]$. By construction, we have that $P'_{(1)} \geq P_{(1)}$ where $P'_{(1)} := \min_{j\in[k]} P'_j$. Hence,
\begin{equation*}
     f_{\mathrm{Sid},k}(P'_1,\dots,P'_k) =  1-(1-P'_{(1)})^k \geq  1-(1-P_{(1)})^k =  f_{\mathrm{Sid},k}(P_1,\dots,P_k),
\end{equation*}
which means that $f_{\mathrm{Sid},k}$ is nondecreasing in each of its arguments.
\end{proof}

\begin{proof}[Proof of \lemref{sidak_combine}$(iii)$]
Suppose that $H^{\star}$ is true. Since each $P_j$ is superuniform under $H^{\star}$, we have that
\begin{equation}\label{sidak_superuniform}
    \Pr(P_j>t)\geq 1-t
    \quad \text{for all $t\in[0,1]$ and $j\in[k]$}.
\end{equation}

Hence, for any $t\in[0,1]$,
\begin{align*}
    \Pr(f_{\mathrm{Sid},k}(P_1,\dots,P_k)\leq t)
    &=
    \Pr(1-(1-P_{(1)})^k\leq t) \\
    &=
    \Pr(P_{(1)}\leq 1-(1-t)^{1/k}) \\
    &=
    1-
    \Pr(P_j>1-(1-t)^{1/k}
    \text{ for all }j\in[k]) \\
    &=
    1-
    \prod_{j\in[k]}
    \Pr(P_j>1-(1-t)^{1/k}) \\
    &\leq
    1-
    \prod_{j\in[k]}(1-t)^{1/k} \\
    &=t,
\end{align*}
where the inequality follows from \eqref{sidak_superuniform}. Therefore, $f_{\mathrm{Sid},k}(P_1,\dots,P_k)$ is a valid $p$-value for testing the intersection null $H^{\star}$.
\end{proof}

\begin{lemma}[Bound on elementary symmetric polynomials]\label{lem:e}
Let $q_1,\dots,q_h,c\in(0,1)$ satisfy $q_j\leq c$ for all $j\in[h]$, and define
\begin{equation*}
    w_j=\frac{q_j}{1-q_j}
    \quad \text{and} \quad
    w=\frac{c}{1-c}.
\end{equation*}
For $s\in\{0,\dots,h\}$, let
\begin{equation*}
    e_s(w_1,\dots,w_h)
    :=
    \sum_{\substack{\mathcal A\subseteq[h]\\|\mathcal A|=s}}
    \prod_{r\in\mathcal A}w_r
\end{equation*}
be the $s$th elementary symmetric polynomial based on $(w_1,\dots,w_h)$, with
$e_0(w_1,\dots,w_h)=1$. Then, for any
$\ell\in\{0,\dots,h\}$ and $j\in\{0,\dots,h-\ell\}$,
\begin{equation*}
    e_{\ell+j}(w_1,\dots,w_h)
    \leq
    e_\ell(w_1,\dots,w_h)
    \binom{h-\ell}{j}w^j.
\end{equation*}
\end{lemma}

\begin{proof}[Proof of \lemref{e}]
Since $q_r\leq c$ for every $r\in[h]$ and $x/(1-x)$ is increasing on
$(0,1)$, we have that
\begin{equation}\label{w_r_w}
    w_r\leq w
    \quad\text{for all }r\in[h].
\end{equation}
For any subset $\mathcal{A} \subseteq[h]$ such that $|\mathcal{A}| = \ell$, we have
\begin{align}
    \sum_{\substack{\mathcal{J} \subseteq[h]\setminus\mathcal{A} \\
                    |\mathcal{J}|=j}} \ 
    \prod_{r\in\mathcal A\cup\mathcal J}w_r
    &=
    \sum_{\substack{\mathcal{J} \subseteq[h]\setminus\mathcal{A} \\ 
                    |\mathcal{J} |=j}} \
    \left( \prod_{r\in\mathcal{J}}w_r \right)  \left( \prod_{r\in\mathcal{A}}w_r \right) \nonumber \\ 
    &=
    \left(\prod_{r\in\mathcal A}w_r\right)
    \sum_{\substack{\mathcal J\subseteq[h]\setminus\mathcal A\\
                    |\mathcal J|=j}} \
    \prod_{r\in\mathcal J}w_r \nonumber \\
    &\leq
    \left(\prod_{r\in\mathcal A}w_r\right)
    \sum_{\substack{\mathcal J\subseteq[h]\setminus\mathcal A\\
                    |\mathcal J|=j}} \
    \prod_{r\in\mathcal J}w \nonumber  \\ 
    &\leq
    \left(\prod_{r\in\mathcal A}w_r\right)
    \sum_{\substack{\mathcal J\subseteq[h]\setminus\mathcal A\\
                    |\mathcal J|=j}} \
    w^j  \nonumber \\
    &\leq
    \left(\prod_{r\in\mathcal A}w_r\right)
    \binom{h-\ell}{j}w^j, \label{e_part1}
\end{align}
where the first inequality is a result of \eqref{w_r_w}. Summing over all subsets $\mathcal A$ such that $|\mathcal{A}| = \ell$ gives
\begin{align}
    \sum_{\substack{\mathcal A\subseteq[h],\,|\mathcal A|=\ell\\
                     \mathcal J\subseteq[h]\setminus\mathcal A,\,|\mathcal J|=j}}
    \prod_{r\in\mathcal A\cup\mathcal J}w_r
    &= \sum_{\substack{\mathcal A\subseteq[h] \\ |\mathcal A|=\ell}} 
        \sum_{\substack{\mathcal{J} \subseteq[h]\setminus\mathcal{A}\\
                    |\mathcal{J}|=j}} \ 
        \prod_{r\in\mathcal A\cup\mathcal J}w_r \nonumber \\
    &\leq  \sum_{\substack{\mathcal A\subseteq[h] \\ |\mathcal A|=\ell}}   
            \left(\prod_{r\in\mathcal A}w_r\right) \binom{h-\ell}{j}w^j \nonumber \\
    &\leq  e_\ell(w_1,\dots,w_h)
    \binom{h-\ell}{j}w^j, \label{e_part2}
\end{align}
where the first inequality stems from \eqref{e_part1}. Moreover, we have that 
\begin{equation}
     \sum_{\substack{\mathcal A\subseteq[h],\,|\mathcal A|=\ell\\
                     \mathcal J\subseteq[h]\setminus\mathcal A,\,|\mathcal J|=j}}
    \prod_{r\in\mathcal A\cup\mathcal J}w_r
    \geq
    \sum_{\substack{\mathcal{S} \subseteq [h] \\
                     |\mathcal{S}| = j + \ell}} 
    \prod_{r\in \mathcal{S} }w_r
    =  e_{\ell+j}(w_1,\dots,w_h). \label{e_part3}
\end{equation}

Thus, combining \eqref{e_part2} and \eqref{e_part3} gives
\begin{equation*}
    e_{\ell+j}(w_1,\dots,w_h)
    \leq
    e_\ell(w_1,\dots,w_h)
    \binom{h-\ell}{j}w^j.
\end{equation*}
\end{proof}

\begin{lemma}[Non-strict conditional validity of \Sidak\ PC $p$-values]
\label{lem:non_strict_validity}
Under the assumptions of \thmref{Sidak_properties}, for every true PC null $H_i^{u/n}$,
\begin{equation*}
    \Pr(\tilde{S}_i \leq  t\mid\tilde{F}_i \leq t)
    \leq t
    \quad \text{for all $t\in[0,1]$},
\end{equation*}
whenever $\Pr(\tilde{F}_i<t)>0$.
\end{lemma}

\begin{proof}[Proof of \lemref{non_strict_validity}]
It suffices to consider $t\in(0,1)$, since the cases $t=0$ and
$t=1$ are immediate. Let
\begin{equation*}
    c_t=1-(1-t)^{1/(n-u+1)},
\end{equation*}
so that
\begin{equation*}
    1-(1-c_t)^{n-u+1}=t.
\end{equation*}
Since the map $x\mapsto 1-(1-x)^{n-u+1}$ is increasing, we have that
\begin{equation}\label{set_equiv}
    \{\tilde S_i\leq t\}
    =
    \{P_{i(u)}\leq c_t\}
    \quad \text{and} \quad
    \{\tilde F_i\leq t\}
    =
    \{P_{i(u-1)}\leq c_t\}.
\end{equation}

Suppose $H^{u/n}_i$ is true and $r\leq u-1$ of the $n$ base null hypotheses for feature $i$ are false. Without loss of generality, let the $r$ corresponding non-null base $p$-values be denoted by $A_{i1},\dots,A_{ir}$.
Let $A$ be the number of these $r$ $p$-values that are at most $c_t$:
\begin{equation*}
    A = \sum_{j=1}^{r} \mathbf{1}\{ A_{ij} \leq c_t \} \leq r.
\end{equation*}
Let $h = n-r$ be the number of true base null hypotheses for feature $i$. Without loss of generality, let the $h$ corresponding null base $p$-values be denoted by $Q_{i1},\dots,Q_{ih}$. Then define $X$ as the number of these $h$ $p$-values that are at most $c_t$:
\begin{equation*}
    X
    =\sum_{j \in [h]} \mathbf{1}\{ \text{$Q_{ij} \leq c_t$} \} \leq h.
\end{equation*}
By validity and mutual independence of the base $p$-values, $X$ is a
sum of independent Bernoulli random variables with success
probabilities
\begin{equation*}
    q_j := \Pr(Q_{ij} \leq c_t) \leq c_t \quad \text{for each $j \in [h]$.}
\end{equation*}
Moreover, $A$ and $X$ are independent.

Fix any $a\in\{0,\dots,r\}$ such that $\Pr(A=a)>0$, and write
\begin{equation*}
    \ell=u-1-a.
\end{equation*}
Conditional on $A=a$, we have
\begin{equation}\label{set_equiv_2}
    \{P_{i(u-1)}\leq c_t\}
    =
    \{ X + a \geq u - 1  \}
    =
    \{X\geq\ell\}
\end{equation}
and
\begin{equation}\label{set_equiv_3}
    \{P_{i(u)}\leq c_t\} 
    = \{ X + a \geq u \} = 
    \{X\geq\ell+1\}.
\end{equation}
Hence, by \eqref{set_equiv}, \eqref{set_equiv_2}, and \eqref{set_equiv_3}, it suffices to prove the lemma by showing that
\begin{equation}\label{show_t_bound}
    \Pr(X\geq\ell+1)\leq t\Pr(X\geq\ell)
\end{equation}
for every $a\in\{0,\dots,r\}$ such that $\Pr(A=a)>0$. Indeed, since $A$ and $X$ are independent, \eqref{show_t_bound} being true implies that
\begin{align}
    \Pr(\tilde{S}_i\leq t) &= 
    \Pr(P_{i(u)} \leq c_t) \nonumber \\
    &=
    \sum_{a=0}^{r}
    \Pr(A=a)\Pr(X\geq u-a) \nonumber \\
    &\leq
    t\sum_{a=0}^{r}
    \Pr(A=a)\Pr(X\geq u-1-a) \nonumber \\
    &= 
   t \Pr(P_{i(u-1)} \leq c_t) \nonumber \\
    &=
    t\Pr(\tilde{F}_i\leq t). \label{AXkey}
\end{align}
Hence, whenever $\Pr(\tilde{F}_i\leq t)>0$, we have
\begin{equation*}
    \Pr(\tilde{S}_i\leq t\mid\tilde{F}_i\leq t)
     =
    \frac{\Pr(\tilde{S}_i\leq t, \tilde{F}_i\leq t)}{\Pr(\tilde{F}_i\leq t)}
    =
    \frac{\Pr(\tilde{S}_i\leq t)}{\Pr(\tilde{F}_i\leq t)}
    \leq t,
\end{equation*}
where the inequality follows from \eqref{AXkey}.

It remains to prove \eqref{show_t_bound}. If $\Pr(X\geq\ell)=0$, then $\Pr(X\geq\ell+1)=0$, so \eqref{show_t_bound} holds trivially. Hence, we consider the case $\Pr(X\geq\ell)>0$.

For $j=1,\dots,h$, let
\begin{equation*}
    w_j=\frac{q_j}{1-q_j}
    \quad\text{and}\quad
    w=\frac{c_t}{1-c_t},
\end{equation*}
so that $w_j\leq w$. For $s\in\{0,\dots,h\}$, let
\begin{equation*}
    e_s(w_1,\dots,w_h)
    :=
    \sum_{\substack{\mathcal A\subseteq[h]\\ |\mathcal A|=s}}
    \prod_{j\in\mathcal A} w_j,
\end{equation*}
with $e_0(w_1,\dots,w_h)=1$. Since $X$ is the sum of independent Bernoulli random variables with
success probabilities $q_1,\dots,q_h$,
\begin{align*}
    \Pr(X=s)
    &=
    \sum_{\substack{\mathcal A\subseteq[h]\\ |\mathcal A|=s}}
    \prod_{j\in\mathcal A}q_j
    \prod_{j\notin\mathcal A}(1-q_j) \\
    &=
     \prod_{j\in [h]}(1-q_j) \sum_{\substack{\mathcal A\subseteq[h]\\ |\mathcal A|=s}} \frac{
    \prod_{j\in\mathcal A}q_j
    \prod_{j\notin\mathcal A}(1-q_j)}{ \prod_{j\in [h]}(1-q_j)} \\
    &=
    \prod_{j\in [h]}(1-q_j)
    \sum_{\substack{\mathcal A\subseteq[h]\\ |\mathcal A|=s}}
    \prod_{j\in\mathcal A}\frac{q_j}{1-q_j} \\
    &=
    \prod_{j\in [h]}(1-q_j) e_s(w_1,\dots,w_h).
\end{align*}
Moreover, for every $j\in\{0,\dots,h-\ell\}$, we have by \lemref{e} that
\begin{equation*}
    e_{\ell+j}(w_1,\dots,w_h)
    \leq
    e_\ell(w_1,\dots,w_h)
    \binom{h-\ell}{j}w^j,
\end{equation*}
where the inequality also holds when some $q_j=0$ by continuity. Therefore,
\begin{align*}
    \Pr(X\geq\ell)
    & = \sum^h_{s = \ell} \Pr(X =  s) \\
    & = \sum^h_{s = \ell} \prod_{k \in [h]}(1-q_k) e_s(w_1,\dots,w_h) \\
    & = \sum^{h - \ell}_{j = 0} \prod_{k \in [h]}(1-q_k) e_{\ell + j}(w_1,\dots,w_h) \\
    & \leq  \sum^{h - \ell}_{j = 0} \prod_{k \in [h]}(1-q_k)  e_\ell(w_1,\dots,w_h)  \binom{h-\ell}{j}w^j \\
    & = \Pr(X=\ell)
    \sum_{j=0}^{h-\ell}
    \binom{h-\ell}{j}w^j \\
    &=
    \Pr(X=\ell)(1+w)^{h-\ell},
\end{align*}
where the last equality stems from the binomial theorem. It follows that
\begin{equation*}
    \Pr(X=\ell\mid X\geq\ell) = \frac{\Pr(X = \ell)}{\Pr(X \geq \ell)}
    \geq
    (1+w)^{-(h-\ell)}
    =
    (1-c_t)^{h-\ell},
\end{equation*}
and hence
\begin{equation*}
    \Pr(X\geq\ell+1\mid X\geq\ell) = 1 - \Pr(X = \ell \mid X \geq \ell)
    \leq
    1-(1-c_t)^{h-\ell}.
\end{equation*}

Finally,
\begin{align*}
    h-\ell
    &=
    (n-r)-(u-1-a) \\
    &=
    n-u+1-(r-a) \\
    &\leq
    n-u+1.
\end{align*}
Thus,
\begin{equation*}
    \Pr(X\geq\ell+1\mid X\geq\ell) \leq
    1-(1-c_t)^{h-\ell}
    \leq
    1-(1-c_t)^{n-u+1}
    =
    t.
\end{equation*}
Multiplying both sides by $\Pr(X\geq\ell)>0$ establishes \eqref{show_t_bound}, completing the proof.
\end{proof}

\begin{lemma}[Characterisation of the AdaFilter-Sid filtering event]
\label{lem:sidak_selection_event}
Let the set of PC hypotheses passing the filter of
AdaFilter-Sid be defined as
\begin{equation*}
    \tilde{\cA}
    :=
    \left\{
        i\in[m]:
        \tilde{F}_i<\hat{t}_{\mathrm{Sid}}
    \right\}.
\end{equation*}
For any fixed $\cA_0\subseteq[m]$, define the auxiliary
threshold
\begin{equation*}
    \hat{t}_{\cA_0}
    :=
    \sup\left\{
        t\in[0,\alpha]:
        1-(1-t)^{
            |\cA_0|
            +
            \sum_{\ell\in[m]\setminus\cA_0}
            \mathbf{1}\{\tilde{F}_\ell<t\}
        }
        \leq\alpha
    \right\}.
\end{equation*}
Then
\begin{align*}
    \{\tilde{\cA}=\cA_0\}
    &=
    \left\{
        \tilde{F}_i<\hat{t}_{\cA_0}
        \text{ for all }i\in\cA_0
    \right\}
    \cap
    \left\{
        \tilde{F}_\ell\geq\hat{t}_{\cA_0}
        \text{ for all }\ell\in[m]\setminus\cA_0
    \right\}.
\end{align*}
\end{lemma}

\begin{proof}[Proof of \lemref{sidak_selection_event}]
Define the feasible region
\begin{align*}
    \cT_{\cA_0}
    :=
    \left\{
        t\in[0,\alpha]:
        1-(1-t)^{|\cA_0|+\tilde{N}_{\cA_0}(t)}
        \leq\alpha
    \right\}
\end{align*}
so that $\hat{t}_{\cA_0}=\sup\cT_{\cA_0}$ by the definition in the lemma.
Since $\tilde{N}(t)$ and $\tilde{N}_{\cA_0}(t)$ are left-continuous, we have
$\hat{t}_{\mathrm{Sid}}\in\cT_{\mathrm{Sid}}$ and
$\hat{t}_{\cA_0}\in\cT_{\cA_0}$.

For every $t\in[0,\alpha]$, we have
\begin{align*}
    \tilde{N}(t)
    =
    \sum_{i\in\cA_0}
    \mathbf{1}\{\tilde{F}_i<t\}
    +
    \tilde{N}_{\cA_0}(t)
    \leq
    |\cA_0|+\tilde{N}_{\cA_0}(t).
\end{align*}
Thus, $\cT_{\cA_0}\subseteq\cT_{\mathrm{Sid}}$, which implies
\begin{equation}\label{threshold_order_A}
    \hat{t}_{\cA_0}
    \leq
    \hat{t}_{\mathrm{Sid}}.
\end{equation}

Suppose that $\tilde{\cA}=\cA_0$.
By the definition of $\tilde{\cA}$, we have that
\begin{equation}\label{tildecAconsequence}
    \tilde{F}_i<\hat{t}_{\mathrm{Sid}}
    \quad\text{for all }i\in\cA_0,
    \qquad
    \tilde{F}_\ell\geq\hat{t}_{\mathrm{Sid}}
    \quad\text{for all }\ell\in[m]\setminus\cA_0.
\end{equation}
Consequently,
\begin{equation*}
    \tilde{N}(\hat{t}_{\mathrm{Sid}})
    =
    |\cA_0|
    \quad\text{and}\quad
    \tilde{N}_{\cA_0}(\hat{t}_{\mathrm{Sid}})=0.
\end{equation*}
Since $\hat{t}_{\mathrm{Sid}}\in\cT_{\mathrm{Sid}}$, we have
\begin{align*}
    \alpha
    &\geq
    1-(1-\hat{t}_{\mathrm{Sid}})^{
        \tilde{N}(\hat{t}_{\mathrm{Sid}})
    }
    =
    1-(1-\hat{t}_{\mathrm{Sid}})^{
        |\cA_0|+\tilde{N}_{\cA_0}(\hat{t}_{\mathrm{Sid}})
    }.
\end{align*}
Hence, $\hat{t}_{\mathrm{Sid}}\in\cT_{\cA_0}$,
which implies
\begin{equation*}
    \hat{t}_{\mathrm{Sid}}
    \leq
    \hat{t}_{\cA_0}.
\end{equation*}
Combining this with \eqref{threshold_order_A}
gives
\begin{equation}\label{LHS_imply_RHS_1}
    \hat{t}_{\cA_0}
    =
    \hat{t}_{\mathrm{Sid}}.
\end{equation}
Thus, \eqref{tildecAconsequence} and
\eqref{LHS_imply_RHS_1} establish that
\begin{align}\label{LHS_imply_RHS}
    \{\tilde{\cA}=\cA_0\}
    \implies
    \left\{
        \tilde{F}_i<\hat{t}_{\cA_0}
        \text{ for all }i\in\cA_0
    \right\}
    \cap
    \left\{
        \tilde{F}_\ell\geq\hat{t}_{\cA_0}
        \text{ for all }\ell\in[m]\setminus\cA_0
    \right\}.
\end{align}

Conversely, suppose that
\begin{equation}\label{sidak_event_B}
    \tilde{F}_i<\hat{t}_{\cA_0}
    \quad\text{for all }i\in\cA_0,
    \qquad
    \tilde{F}_\ell\geq\hat{t}_{\cA_0}
    \quad\text{for all }\ell\in[m]\setminus\cA_0.
\end{equation}
The function
$t\mapsto1-(1-t)^{|\cA_0|+\tilde{N}_{\cA_0}(t)}$
is nondecreasing on $[0,\alpha]$. Consequently,
$\cT_{\cA_0}$ is an initial interval, and
for every $t\notin\cT_{\cA_0}$, equivalently
$t\in(\hat{t}_{\cA_0},\alpha]$, we have
\begin{align*}
    \tilde{N}(t)
    =
    \sum_{i\in\cA_0}
    \mathbf{1}\{\tilde{F}_i<t\}
    +
    \tilde{N}_{\cA_0}(t)
    =
    |\cA_0|+\tilde{N}_{\cA_0}(t).
\end{align*}
Since $\hat{t}_{\cA_0}=\sup\cT_{\cA_0}$,
every such $t$ satisfies
\begin{equation*}
    1-(1-t)^{|\cA_0|+\tilde{N}_{\cA_0}(t)}
    >
    \alpha.
\end{equation*}
Therefore,
\begin{equation*}
    1-(1-t)^{\tilde{N}_{\varnothing}(t)}
    >
    \alpha,
\end{equation*}
so that $t\notin\cT_{\mathrm{Sid}}$ for every $t \notin \cT_{\cA_0}$. This therefore implies that $t \in \cT_{\cA_0}$ for every $t \in \cT_{\mathrm{Sid}}$.
It follows then from the formulation of $\hat{t}_{\cA_0}$ that
\begin{equation*}
    \hat{t}_{\mathrm{Sid}}
    \leq
    \hat{t}_{\cA_0}.
\end{equation*}
Together with \eqref{threshold_order_A},
this establishes
\begin{equation*}
    \hat{t}_{\mathrm{Sid}}
    =
    \hat{t}_{\cA_0}.
\end{equation*}
Hence, by \eqref{sidak_event_B},
\begin{equation*}
    \tilde{\cA}
    =
    \left\{
        i\in[m]:
        \tilde{F}_i<\hat{t}_{\mathrm{Sid}}
    \right\}
    =
    \cA_0.
\end{equation*}
This establishes the converse implication and
completes the proof.
\end{proof}

\section{Proof of \lemref{uniform_improvement_AdaFilter} }
\label{app:uniform_improvement_AdaFilter}

\begin{proof}[Proof of \lemref{uniform_improvement_AdaFilter}$(i)$]
For $t\in[0,\alpha]$, define
\begin{equation*}
    N^{(g)}(t)
    :=
    \sum_{i\in[m]}
    \mathbf{1}\{g(F_i)<t\}.
\end{equation*}
Let $\cT_{\mathrm{Bon}}^{(g)}$ denote the set of feasible thresholds for the AdaFilter-Bon procedure based on $\{(g(S_i),g(F_i))\}_{i\in[m]}$, that is,
\begin{equation*}
    \cT_{\mathrm{Bon}}^{(g)}
    :=
    \left\{
        t\in[0,\alpha]:
        tN^{(g)}(t)\leq\alpha
    \right\}.
\end{equation*}
Moreover, let the corresponding rejection threshold be
\begin{equation}\label{t_hat_g}
    \hat{t}_{\mathrm{Bon}}^{(g)}
    :=
    \sup\cT_{\mathrm{Bon}}^{(g)}.
\end{equation}

Since $g$ is strictly increasing, we have for every $t\in[0,\alpha]$ that
\begin{equation}\label{filter_count_transform}
    N^{(g)}(g(t))
    =
    \sum_{i\in[m]}
    \mathbf{1}\{g(F_i)<g(t)\}
    =
    \sum_{i\in[m]}
    \mathbf{1}\{F_i<g^{-1}(g(t))\}
    =
    N(t).
\end{equation}
Moreover, since $0\leq g(t)\leq t$, any $t$ satisfying
\begin{equation*}
    tN(t)\leq\alpha
\end{equation*}
also satisfies
\begin{align*}
    g(t)N^{(g)}(g(t))
    =
    g(t)N(t)
    \leq
    tN(t)
    \leq\alpha,
\end{align*}
where the equality follows from \eqref{filter_count_transform}. Thus, if $t\in\cT_{\mathrm{Bon}}$, then it must follow that $g(t)\in\cT_{\mathrm{Bon}}^{(g)}$. Consequently, by the formulation of $\hat{t}_{\mathrm{Bon}}^{(g)}$ given in \eqref{t_hat_g}, we have
\begin{equation}\label{t_hat_g_res}
    \hat{t}_{\mathrm{Bon}}^{(g)}
    \geq g(t)
    \quad\text{for all }t\in\cT_{\mathrm{Bon}}.
\end{equation}
Since $N(t)$ is left-continuous, we have that $\hat{t}_{\mathrm{Bon}}\in\cT_{\mathrm{Bon}}$, which then means by \eqref{t_hat_g_res} that
\begin{equation}\label{t_hat_g_inequality}
    \hat{t}_{\mathrm{Bon}}^{(g)}
    \geq
    g(\hat{t}_{\mathrm{Bon}}).
\end{equation}
Hence, it follows that
\begin{align*}
    \cR_{\mathrm{Bon}}
    &=
    \{i\in[m]:S_i<\hat{t}_{\mathrm{Bon}}\}\\
    &=
    \{i\in[m]:g(S_i)<g(\hat{t}_{\mathrm{Bon}})\}\\
    &\subseteq
    \{i\in[m]:g(S_i)<\hat{t}_{\mathrm{Bon}}^{(g)}\}\\
    &=
    \cR_{\mathrm{Bon}}^{(g)},
\end{align*}
where the inclusion follows from \eqref{t_hat_g_inequality}.

Since the base $p$-values in $(P_{ij})_{(i,j)\in[m]\times[n]}$ are mutually independent, and each $g(S_i)$ is a valid PC $p$-value satisfying the conditional validity property given $g(F_i)$, with $g(F_i)\leq g(S_i)$ by the strict monotonicity of $g$, the FWER argument for AdaFilter-Bon applies directly to $\{(g(S_i),g(F_i))\}_{i\in[m]}$; see Section S3.3 of \citet{Wang2022}. Hence, the resulting AdaFilter-Bon procedure based on $\{(g(S_i),g(F_i))\}_{i\in[m]}$ controls the FWER at level $\alpha$.
\end{proof}

\begin{proof}[Proof of \lemref{uniform_improvement_AdaFilter}$(ii)$]
For any subset $\cR\subseteq[m]$, define the auxiliary
feasible region
\begin{equation*}
    \cT_{\cR}
    :=
    \left\{
        t\in[0,\alpha]:
        tN_{\cR}(t)\leq\alpha
    \right\}
\end{equation*}
and the corresponding threshold
$\hat{t}_{\cR}:=\sup\cT_{\cR}$.
For the transformed statistics, define
\begin{equation*}
    N_{\cR}^{(g)}(t)
    :=
    \sum_{i\in[m]\setminus\cR}
    \mathbf{1}\{g(F_i)<t\},
\end{equation*}
the feasible region
\begin{equation*}
    \cT_{\cR}^{(g)}
    :=
    \left\{
        t\in[0,\alpha]:
        tN_{\cR}^{(g)}(t)\leq\alpha
    \right\},
\end{equation*}
and the corresponding threshold
$\hat{t}_{\cR}^{(g)}:=\sup\cT_{\cR}^{(g)}$.

Consider any two subsets
$\cR\subseteq\cR^{(g)}\subseteq[m]$.
Since $g$ is strictly increasing, we have for
every $t\in[0,\alpha]$ that
\begin{align}
    N_{\cR^{(g)}}^{(g)}(g(t))
    &=
    \sum_{i\in[m]\setminus\cR^{(g)}}
    \mathbf{1}\{g(F_i)<g(t)\}
    \nonumber\\
    &=
    \sum_{i\in[m]\setminus\cR^{(g)}}
    \mathbf{1}\{F_i<t\}
    \nonumber\\
    &\leq
    \sum_{i\in[m]\setminus\cR}
    \mathbf{1}\{F_i<t\}
    \nonumber\\
    &=
    N_{\cR}(t).
    \label{ngineq}
\end{align}
Since $g(0)=0$, $g$ is strictly increasing,
and $g(t)\leq t$, we have
$0\leq g(t)\leq t\leq\alpha$ for every
$t\in[0,\alpha]$.
Consequently, whenever
$tN_{\cR}(t)\leq\alpha$, i.e.,
$t\in\cT_{\cR}$, we have
\begin{align*}
    g(t)N_{\cR^{(g)}}^{(g)}(g(t))
    &\leq
    g(t)N_{\cR}(t)\\
    &\leq
    tN_{\cR}(t)\\
    &\leq\alpha,
\end{align*}
where the first inequality follows from
\eqref{ngineq}.
Thus,
$g(t)\in\cT_{\cR^{(g)}}^{(g)}$.

Since $N_{\cR}(t)$ is left-continuous,
we have $\hat{t}_{\cR}\in\cT_{\cR}$,
and it follows that
$g(\hat{t}_{\cR})\in
\cT_{\cR^{(g)}}^{(g)}$.
Therefore, by the definition of
$\hat{t}_{\cR^{(g)}}^{(g)}$, we obtain
\begin{equation}\label{seqbon_threshold_inequality}
    \hat{t}_{\cR^{(g)}}^{(g)}
    \geq
    g(\hat{t}_{\cR})
    \quad\text{whenever }
    \cR\subseteq\cR^{(g)}.
\end{equation}

We now proceed inductively over the sequential
iterations. Let $\cR^{(k)}$ and
$\cR^{(k,g)}$ denote the rejection sets after
iteration $k$ of AdaFilter-SeqBon and its
variant based on
$\{(g(S_i),g(F_i))\}_{i\in[m]}$,
respectively, with
$\cR^{(0)}=\cR^{(0,g)}=\varnothing$.
If either procedure terminates, we continue
to apply its update rule, which leaves its
rejection set unchanged.

At the first iteration,
\eqref{seqbon_threshold_inequality} implies that
\begin{equation}\label{hattgR0}
    \hat{t}_{\cR^{(0,g)}}^{(g)}
    \geq
    g(\hat{t}_{\cR^{(0)}}),
\end{equation}
so every hypothesis rejected by AdaFilter-SeqBon
at the first iteration is also rejected by its
variant based on
$\{(g(S_i),g(F_i))\}_{i\in[m]}$.
Hence,
\begin{align*}
    \cR^{(1)}
    &=
    \cR^{(0)}
    \cup
    \left\{
        i\notin\cR^{(0)}:
        S_i<\hat{t}_{\cR^{(0)}}
    \right\}\\
    &=
    \cR^{(0,g)}
    \cup
    \left\{
        i\notin\cR^{(0,g)}:
        S_i<\hat{t}_{\cR^{(0)}}
    \right\}\\
    &=
    \cR^{(0,g)}
    \cup
    \left\{
        i\notin\cR^{(0,g)}:
        g(S_i)<g(\hat{t}_{\cR^{(0)}})
    \right\}\\
    &\subseteq
    \cR^{(0,g)}
    \cup
    \left\{
        i\notin\cR^{(0,g)}:
        g(S_i)<\hat{t}_{\cR^{(0,g)}}^{(g)}
    \right\}\\
    &=
    \cR^{(1,g)},
\end{align*}
where the inclusion follows from
\eqref{hattgR0}.

Suppose that
$\cR^{(k-1)}\subseteq\cR^{(k-1,g)}$
for some $k\geq2$.
By \eqref{seqbon_threshold_inequality},
we have
\begin{equation}\label{hattgk1}
    \hat{t}_{\cR^{(k-1,g)}}^{(g)}
    \geq
    g(\hat{t}_{\cR^{(k-1)}}).
\end{equation}
For any hypothesis $i$ newly rejected by
AdaFilter-SeqBon at iteration $k$, we have
\begin{equation*}
    i\notin\cR^{(k-1)}
    \quad\text{and}\quad
    S_i<\hat{t}_{\cR^{(k-1)}}.
\end{equation*}
Since $g$ is strictly increasing, this implies
\begin{equation*}
    g(S_i)
    <
    g(\hat{t}_{\cR^{(k-1)}})
    \leq
    \hat{t}_{\cR^{(k-1,g)}}^{(g)},
\end{equation*}
where the last inequality follows from
\eqref{hattgk1}.
Thus, $i$ is either already contained in
$\cR^{(k-1,g)}$ or is newly rejected by
the AdaFilter-SeqBon variant based on
$\{(g(S_i),g(F_i))\}_{i\in[m]}$
at iteration $k$.
Consequently,
$\cR^{(k)}\subseteq\cR^{(k,g)}$.
By induction, the inclusion holds at every
iteration, yielding
\begin{equation*}
    \cR_{\mathrm{SeqBon}}
    \subseteq
    \cR_{\mathrm{SeqBon}}^{(g)}.
\end{equation*}

It remains to establish FWER control.
Define the following rejection threshold based
on $\cH^{u/n}_0$, the set of true PC nulls:
\begin{equation*}
    \hat{t}^{(g)}(\cH^{u/n}_0)
    :=
    \sup\left\{
        t\in[0,\alpha]:
        t\sum_{i\in\cH^{u/n}_0}
        \mathbf{1}\{g(F_i)<t\}
        \leq\alpha
    \right\}.
\end{equation*}
Suppose that the AdaFilter-SeqBon procedure
based on
$\{(g(S_i),g(F_i))\}_{i\in[m]}$
makes at least one false rejection, and let
$k$ be the first iteration at which this occurs.
Then
$\cR^{(k-1,g)}\cap\cH^{u/n}_0=\varnothing$,
which means
$\cH^{u/n}_0\subseteq
[m]\setminus\cR^{(k-1,g)}$.
Therefore,
\begin{align*}
    N_{\cR^{(k-1,g)}}^{(g)}(t)
    &=
    \sum_{i\in[m]\setminus\cR^{(k-1,g)}}
    \mathbf{1}\{g(F_i)<t\}\\
    &\geq
    \sum_{i\in\cH^{u/n}_0}
    \mathbf{1}\{g(F_i)<t\}
\end{align*}
for all $t\in[0,\alpha]$.
This implies that
\begin{equation*}
    \hat{t}_{\cR^{(k-1,g)}}^{(g)}
    \leq
    \hat{t}^{(g)}(\cH^{u/n}_0).
\end{equation*}
Since at least one true PC null is rejected
at iteration $k$, there exists some
$i\in\cH^{u/n}_0$ such that
\begin{equation*}
    g(S_i)
    <
    \hat{t}_{\cR^{(k-1,g)}}^{(g)}
    \leq
    \hat{t}^{(g)}(\cH^{u/n}_0).
\end{equation*}
Consequently,
\begin{align*}
    &\Pr\left(
        \cR_{\mathrm{SeqBon}}^{(g)}
        \cap\cH^{u/n}_0\neq\varnothing
    \right)\\
    &\quad\leq
    \Pr\left(
        g(S_i)<\hat{t}^{(g)}(\cH^{u/n}_0)
        \text{ for some }i\in\cH^{u/n}_0
    \right)\\
    &\quad\leq\alpha,
\end{align*}
where the last inequality follows from the
FWER control of the one-step AdaFilter
procedure based on
$\{(g(S_i),g(F_i))\}_{i\in\cH^{u/n}_0}$,
as established in part $(i)$, since its
assumptions continue to hold upon restriction
to the subfamily of true PC nulls.

Hence, the sequential procedure based on
$\{(g(S_i),g(F_i))\}_{i\in[m]}$
controls the FWER at level $\alpha$.
\end{proof}

\section{Proof of \thmref{Sidak_properties} }
\label{app:Sidak_properties}

\begin{proof}[Proof of \thmref{Sidak_properties}$(i)$]
By \lemref{sidak_combine}, the \Sidak\ combining function $f_{\mathrm{Sid},k}$ is symmetric, coordinatewise nondecreasing, and valid for testing the intersection of $k$ null hypotheses. Hence, Theorem 2.1 of \citet{Wang2019}, with $k=n-u+1$, implies that
\begin{align*}
    f_{\mathrm{Sid},n-u+1}
    (P_{i(u)},P_{i(u+1)},\dots,P_{i(n)})
    &=
    1-(1-P_{i(u)})^{n-u+1}
\end{align*}
is a valid PC $p$-value for testing $H_i^{u/n}$. This is precisely $\tilde{S}_i$, and hence $\tilde{S}_i$ is a valid PC $p$-value for testing $H_i^{u/n}$.
\end{proof}

\begin{proof}[Proof of \thmref{Sidak_properties}$(ii)$]
The case $t=0$ is vacuous, so consider any $t\in(0,1]$
such that $\Pr(\tilde{F}_i<t)>0$. By \lemref{non_strict_validity} and the fact that
$\tilde{F}_i\leq\tilde{S}_i$, for every $s<t$
sufficiently close to $t$, we have
\begin{align*}
    s
    \geq
    \Pr(\tilde{S}_i\leq s\mid\tilde{F}_i\leq s)
    =
    \frac{\Pr(\tilde{S}_i\leq s,\tilde{F}_i\leq s)}
    {\Pr(\tilde{F}_i\leq s)}
    =
    \frac{\Pr(\tilde{S}_i\leq s)}
    {\Pr(\tilde{F}_i\leq s)}.
\end{align*}
Consequently,
\begin{equation}\label{eq:sidak_strict_step}
    \Pr(\tilde{S}_i\leq s)
    \leq
    s\Pr(\tilde{F}_i\leq s).
\end{equation}

Now, let $(s_k)_{k\geq 1}$ be a sequence satisfying
$s_k<t$ and $s_k \uparrow t$. Since
\begin{equation*}
    \{\tilde{S}_i<t\}
    =
    \bigcup_{k=1}^{\infty}
    \{\tilde{S}_i\leq s_k\},
\end{equation*}
and the events on the right-hand side form an
increasing sequence, continuity of probability from
below gives
\begin{equation*}
    \lim_{k\to\infty}
    \Pr(\tilde{S}_i\leq s_k)
    =
    \Pr(\tilde{S}_i<t).
\end{equation*}
Similarly,
\begin{equation*}
    \lim_{k\to\infty}
    \Pr(\tilde{F}_i\leq s_k)
    =
    \Pr(\tilde{F}_i<t).
\end{equation*}
Therefore, letting $k\to\infty$ in
\eqref{eq:sidak_strict_step} yields
\begin{equation}\label{cp}
    \Pr(\tilde{S}_i<t)
    \leq
    t\Pr(\tilde{F}_i<t).
\end{equation}

Finally, since $\tilde{F}_i\leq\tilde{S}_i$,
we have that
\begin{align*}
    \Pr(\tilde{S}_i<t\mid\tilde{F}_i<t)
    =
    \frac{\Pr(\tilde{S}_i<t,\tilde{F}_i<t)}
    {\Pr(\tilde{F}_i<t)}
    =
    \frac{\Pr(\tilde{S}_i<t)}
    {\Pr(\tilde{F}_i<t)}
    \leq t,
\end{align*}
where the inequality follows from \eqref{cp}.
\end{proof}

\section{Proof of \thmref{sidak_g_optimal} }
\label{app:sidak_g_optimal}

\begin{proof}
Let $g:[0,n-u+1]\to[0,1]$ be continuous and
strictly increasing, with $g(0)=0$ and
$g(x)\leq x$ for all $x\in[0,n-u+1]$.
We first show that $g_{\mathrm{Sid}}(x)\leq g(x)$
for all $x\in[0,n-u+1]$ implies $g\in\cG$.
Since $g_{\mathrm{Sid}}(n-u+1)=1$ and $g$ takes
values in $[0,1]$, we have $g(n-u+1)=1$.
Thus, by continuity and strict monotonicity,
$g^{-1}$ is well defined on $[0,1]$.

Fix $t\in[0,1]$ and let $x=g^{-1}(t)$.
Since $g$ is strictly increasing,
\begin{equation*}
    \{g(S_i)\leq t\}
    =
    \{S_i\leq x\}
    =
    \{g_{\mathrm{Sid}}(S_i)\leq g_{\mathrm{Sid}}(x)\}
    =
    \{\tilde{S}_i\leq g_{\mathrm{Sid}}(x)\}.
\end{equation*}
Therefore, by \thmref{Sidak_properties}$(i)$,
\begin{align*}
    \Pr(g(S_i)\leq t)
    &=
    \Pr(\tilde{S}_i\leq g_{\mathrm{Sid}}(x))\\
    &\leq g_{\mathrm{Sid}}(x)\\
    &\leq g(x)\\
    &=t,
\end{align*}
which establishes the validity of $g(S_i)$.

Similarly, for every $t\in(0,1]$ such that
$\Pr(g(F_i)<t)>0$, by
\thmref{Sidak_properties}$(ii)$, we have
\begin{align*}
    \Pr(g(S_i)<t\mid g(F_i)<t)
    &=
    \Pr(S_i<x\mid F_i<x)\\
    &=
    \Pr\bigl(
        \tilde{S}_i<g_{\mathrm{Sid}}(x)
        \mid
        \tilde{F}_i<g_{\mathrm{Sid}}(x)
    \bigr)\\
    &\leq g_{\mathrm{Sid}}(x)\\
    &\leq g(x)\\
    &=t.
\end{align*}
Hence, $g(S_i)$ is also conditionally valid
given $g(F_i)$, establishing that $g\in\cG$.

Conversely, suppose that $g\in\cG$. Consider a
configuration under $H_i^{u/n}$ in which exactly
$u-1$ base hypotheses are false, with their
corresponding $p$-values identically equal to zero,
while the remaining $n-u+1$ base $p$-values are
mutually independent $\operatorname{Unif}(0,1)$
random variables $U_1,\dots,U_{n-u+1}$.
Under this configuration,
\begin{equation*}
    P_{i(u)}
    =
    \min_{j\in[n-u+1]}U_j,
\end{equation*}
and hence
\begin{equation*}
    S_i
    =
    (n-u+1)\min_{j\in[n-u+1]}U_j.
\end{equation*}
For any $x\in[0,n-u+1]$, strict monotonicity
of $g$ gives
\begin{align*}
    \Pr(g(S_i)\leq g(x))
    &=
    \Pr(S_i\leq x)\\
    &=
    \Pr\left(
        \min_{j\in[n-u+1]}U_j
        \leq\frac{x}{n-u+1}
    \right)\\
    &=
    1-\left(1-\frac{x}{n-u+1}\right)^{n-u+1}\\
    &=
    g_{\mathrm{Sid}}(x).
\end{align*}
Since $g(S_i)$ is assumed to be a valid PC
$p$-value for testing $H_i^{u/n}$, it follows that
\begin{equation*}
    g_{\mathrm{Sid}}(x)
    =
    \Pr(g(S_i)\leq g(x))
    \leq g(x)
\end{equation*}
for every $x\in[0,n-u+1]$.
This establishes the converse implication.

Finally, $g_{\mathrm{Sid}}$ is continuous and strictly
increasing, satisfies $g_{\mathrm{Sid}}(0)=0$, and
obeys $g_{\mathrm{Sid}}(x)\leq x$ for all
$x\in[0,n-u+1]$. Taking $g=g_{\mathrm{Sid}}$ in
the first part of the proof therefore shows that
$g_{\mathrm{Sid}}\in\cG$. Since every $g\in\cG$
satisfies $g_{\mathrm{Sid}}(x)\leq g(x)$ for all
$x\in[0,n-u+1]$ by the second part of the proof,
we conclude that $g_{\mathrm{Sid}}$ is the pointwise
smallest transformation in $\cG$.
\end{proof}

\section{Proof of \corref{sidak_rejection_optimal}}
\begin{proof}[\unskip\nopunct]
Fix any $g\in\cG$.
By \thmref{sidak_g_optimal},
\begin{equation*}
    g_{\mathrm{Sid}}(n-u+1)=g(n-u+1)=1.
\end{equation*}
Since $g$ is continuous and strictly increasing,
its inverse is well defined on $[0,1]$.

Define
\begin{equation*}
    h(t):=
    g_{\mathrm{Sid}}(g^{-1}(t)),
    \qquad t\in[0,1].
\end{equation*}
By \thmref{sidak_g_optimal}, $h$ is
continuous and strictly increasing, with
$h(0)=0$ and $h(t)\leq t$ for all
$t\in[0,1]$.
Moreover,
\begin{equation*}
    h(g(S_i))=g_{\mathrm{Sid}}(S_i),
    \qquad
    h(g(F_i))=g_{\mathrm{Sid}}(F_i).
\end{equation*}

Consequently, the dominance arguments in
the proofs of
\lemref{uniform_improvement_AdaFilter}$(i)$
and $(ii)$ apply with the original statistics
$(S_i,F_i)$ replaced by
$(g(S_i),g(F_i))$ and the transformation
$g$ replaced by $h$.
This yields
\begin{equation*}
    \cR_{\mathrm{Bon}}^{(g)}
    \subseteq
    \cR_{\mathrm{Bon}}^{(g_{\mathrm{Sid}})}
    \qquad\text{and}\qquad
    \cR_{\mathrm{SeqBon}}^{(g)}
    \subseteq
    \cR_{\mathrm{SeqBon}}^{(g_{\mathrm{Sid}})}.
\end{equation*}
\end{proof}

\section{Proof of \thmref{sidak_fwer} }
\label{app:sidak_fwer}

\begin{proof}[Proof of \thmref{sidak_fwer}$(i)$]
By \thmref{Sidak_properties}$(ii)$, for every true PC null $H_i^{u/n}$,
\begin{equation*}
    \Pr(\tilde{S}_i<t\mid\tilde{F}_i<t)
    \leq t
    \quad\text{for all }t\in(0,1],
\end{equation*}
whenever $\Pr(\tilde{F}_i<t)>0$.
Consequently,
\begin{equation}\label{sidak_joint_bound}
    \Pr(\tilde{S}_i\geq t,\tilde{F}_i<t)
    \geq (1-t)\Pr(\tilde{F}_i<t)
    \quad\text{for all }t\in[0,1].
\end{equation}
Let $\tilde{\cA}$ be the set of features passing the filter
of AdaFilter-Sid, as defined in \lemref{sidak_selection_event}.
For any fixed $\cA_0\subseteq[m]$, let $\hat{t}_{\cA_0}$
be the auxiliary threshold defined in the same lemma.
Let $\mathcal{F}_{-\cA_0}$ denote the $\sigma$-algebra generated by the base $p$-values from features outside $\cA_0$, i.e.,
\begin{equation*}
    \mathcal{F}_{-\cA_0}
    :=
    \sigma\left(
        P_{\ell j}:
        \ell\in[m]\setminus\cA_0,\ j\in[n]
    \right).
\end{equation*}
By construction, $\hat{t}_{\cA_0}$ is measurable with respect to $\mathcal{F}_{-\cA_0}$. Hence, under mutual independence of all the base $p$-values in
$\{P_{ij}\}_{(i,j)\in[m]\times[n]}$, we have that
$\mathcal{F}_{-\cA_0}$ is independent of
$\{(\tilde{S}_i,\tilde{F}_i):i\in\cA_0\}$.

Since $\tilde{F}_i\leq\tilde{S}_i$ for all $i\in[m]$, features outside $\tilde{\cA}$
cannot be rejected. Hence, using
\lemref{sidak_selection_event} and mutual independence
across features, we have

\begin{align}
    &\Pr\left(
        \cR_{\mathrm{Sid}}\cap\cH^{u/n}_0=\varnothing,
        \tilde{\cA}=\cA_0
        \,\middle|\,
        \mathcal{F}_{-\cA_0}
    \right)
    \nonumber\\
    &=
    \Pr\Big(
        \tilde{S}_i\geq\hat{t}_{\cA_0}
        \text{ for all }i\in\cA_0\cap\cH^{u/n}_0,
        \nonumber\\
    &\qquad\qquad
        \tilde{F}_i<\hat{t}_{\cA_0}
        \text{ for all }i\in\cA_0,
        \nonumber\\
    &\qquad\qquad
        \tilde{F}_\ell\geq\hat{t}_{\cA_0}
        \text{ for all }\ell\in[m]\setminus\cA_0
        \,\mid\,
        \mathcal{F}_{-\cA_0}
    \Big)
    \nonumber\\
    &=
    \mathbf{1}\left\{
        \tilde{F}_\ell\geq\hat{t}_{\cA_0}
        \text{ for all }\ell\in[m]\setminus\cA_0
    \right\}
    \nonumber\\
    &\quad\times
    \prod_{i\in\cA_0\cap\cH^{u/n}_0}
    \Pr\left(
        \tilde{S}_i\geq\hat{t}_{\cA_0},
        \tilde{F}_i<\hat{t}_{\cA_0}
        \,\middle|\,
        \mathcal{F}_{-\cA_0}
    \right)
    \nonumber\\
    &\quad\times
    \prod_{i\in\cA_0\setminus\cH^{u/n}_0}
    \Pr\left(
        \tilde{F}_i<\hat{t}_{\cA_0}
        \,\middle|\,
        \mathcal{F}_{-\cA_0}
    \right)
    \nonumber\\
    &\geq
    (1-\hat{t}_{\cA_0})^{|\cA_0\cap\cH^{u/n}_0|}
    \mathbf{1}\left\{
        \tilde{F}_\ell\geq\hat{t}_{\cA_0}
        \text{ for all }\ell\in[m]\setminus\cA_0
    \right\}
    \nonumber\\
    &\quad\times
    \prod_{i\in\cA_0}
    \Pr\left(
        \tilde{F}_i<\hat{t}_{\cA_0}
        \,\middle|\,
        \mathcal{F}_{-\cA_0}
    \right)
    \nonumber\\
    &=
    (1-\hat{t}_{\cA_0})^{|\cA_0\cap\cH^{u/n}_0|}
    \Pr\left(
        \tilde{\cA}=\cA_0
        \,\middle|\,
        \mathcal{F}_{-\cA_0}
    \right),
    \label{sidak_reject_ineq}
\end{align}

where the inequality follows from
\eqref{sidak_joint_bound}, since
$\hat{t}_{\cA_0}$ is measurable with respect
to $\mathcal{F}_{-\cA_0}$ and independent of
the statistics from features in $\cA_0$.

On the event
$\{\tilde{F}_\ell\geq\hat{t}_{\cA_0}
\text{ for all }\ell\notin\cA_0\}$,
the definition of $\hat{t}_{\cA_0}$ gives
\begin{equation*}
    1-(1-\hat{t}_{\cA_0})^{|\cA_0|}
    \leq\alpha.
\end{equation*}
Consequently,
\begin{equation*}
    (1-\hat{t}_{\cA_0})^{|\cA_0\cap\cH^{u/n}_0|}
    \geq
    (1-\hat{t}_{\cA_0})^{|\cA_0|}
    \geq 1-\alpha.
\end{equation*}
Combining this with \eqref{sidak_reject_ineq},
and noting that
$\Pr(\tilde{\cA}=\cA_0\mid\mathcal{F}_{-\cA_0})=0$
when the above event does not hold, yields
\begin{align*}
    &\Pr\left(
        \cR_{\mathrm{Sid}}\cap\cH^{u/n}_0=\varnothing,
        \tilde{\cA}=\cA_0
        \,\middle|\,
        \mathcal{F}_{-\cA_0}
    \right)\\
    &\qquad\geq
    (1-\alpha)
    \Pr\left(
        \tilde{\cA}=\cA_0
        \,\middle|\,
        \mathcal{F}_{-\cA_0}
    \right).
\end{align*}
Taking expectations gives
\begin{equation*}
    \Pr\left(
        \cR_{\mathrm{Sid}}\cap\cH^{u/n}_0=\varnothing,
        \tilde{\cA}=\cA_0
    \right)
    \geq
    (1-\alpha)\Pr(\tilde{\cA}=\cA_0).
\end{equation*}
Thus,
\begin{align*}
    \Pr\left(
        \cR_{\mathrm{Sid}}\cap\cH^{u/n}_0\neq\varnothing
    \right)
    &=
    1-\sum_{\cA_0\subseteq[m]}
    \Pr\left(
        \cR_{\mathrm{Sid}}\cap\cH^{u/n}_0=\varnothing,
        \tilde{\cA}=\cA_0
    \right)\\
    &\leq
    1-(1-\alpha)
    \sum_{\cA_0\subseteq[m]}
    \Pr(\tilde{\cA}=\cA_0)\\
    &=\alpha.
\end{align*}
This establishes the FWER control of AdaFilter-Sid.

We next establish that
$\tilde{\cR}_{\mathrm{Bon}}\subseteq\cR_{\mathrm{Sid}}$.
Let $\tilde{t}_{\mathrm{Bon}}$ denote the AdaFilter-Bon rejection threshold obtained using
$\{(\tilde{S}_i,\tilde{F}_i)\}_{i\in[m]}$, so that
\begin{equation*}
    \tilde{t}_{\mathrm{Bon}}
    :=
    \sup\widetilde{\cT}_{\mathrm{Bon}},
\end{equation*}
where
\begin{equation*}
    \widetilde{\cT}_{\mathrm{Bon}}
    :=
    \left\{
        t\in[0,\alpha]:
        t\tilde{N}(t)\leq\alpha
    \right\}.
\end{equation*}
For every nonnegative integer $N$, Bernoulli's inequality gives
\begin{equation}\label{eq:sidak_vs_bon_mult}
    1-(1-t)^N
    \leq Nt.
\end{equation}
Hence, every $t\in\widetilde{\cT}_{\mathrm{Bon}}$
also satisfies $t\in\cT_{\mathrm{Sid}}$, i.e.,
$\widetilde{\cT}_{\mathrm{Bon}}\subseteq\cT_{\mathrm{Sid}}$.
It follows that
\begin{equation*}
    \tilde{t}_{\mathrm{Bon}}
    =
    \sup\widetilde{\cT}_{\mathrm{Bon}}
    \leq
    \sup\cT_{\mathrm{Sid}}
    =
    \hat{t}_{\mathrm{Sid}},
\end{equation*}
and therefore
\begin{equation*}
    \tilde{\cR}_{\mathrm{Bon}}
    =
    \left\{
        i\in[m]:
        \tilde{S}_i<\tilde{t}_{\mathrm{Bon}}
    \right\}
    \subseteq
    \left\{
        i\in[m]:
        \tilde{S}_i<\hat{t}_{\mathrm{Sid}}
    \right\}
    =
    \cR_{\mathrm{Sid}}.
\end{equation*}
\end{proof}

\begin{proof}[Proof of \thmref{sidak_fwer}$(ii)$]
For any subset $\cR\subseteq[m]$, define
\begin{equation*}
    \hat{t}_{\mathrm{Sid}}(\cR)
    :=
    \sup\left\{
        t\in[0,\alpha]:
        1-(1-t)^{\tilde{N}_{\cR}(t)}
        \leq\alpha
    \right\}.
\end{equation*}
Define the successor function
\begin{equation*}
    \mathcal{D}(\cR)
    :=
    \left\{
        i\in[m]\setminus\cR:
        \tilde{S}_i<\hat{t}_{\mathrm{Sid}}(\cR)
    \right\}.
\end{equation*}
By the sequential rejection principle of
\citet[Theorem~1]{Goeman2010}, the sequential
procedure based on $\mathcal{D}$ controls the
FWER at level $\alpha$ if the following two
conditions hold:
\begin{enumerate}
    \item For every $\cR\subseteq\cR'\subseteq[m]$,
    \begin{equation}\label{sidak_seq_monotonicity}
        \mathcal{D}(\cR)
        \subseteq
        \mathcal{D}(\cR')\cup\cR'.
    \end{equation}

    \item For every configuration of true PC nulls,
    \begin{equation}\label{sidak_seq_single_step}
        \Pr\left(
            \mathcal{D}([m]\setminus\cH^{u/n}_0)
            \cap\cH^{u/n}_0
            \neq\varnothing
        \right)
        \leq\alpha.
    \end{equation}
\end{enumerate}

We first verify \eqref{sidak_seq_monotonicity}.
If $\cR\subseteq\cR'$ for some
$\cR'\subseteq[m]$, then
\begin{equation*}
    \tilde{N}_{\cR'}(t)
    \leq
    \tilde{N}_{\cR}(t)
    \qquad\text{for all }t\in[0,\alpha],
\end{equation*}
and hence
\begin{equation*}
    \hat{t}_{\mathrm{Sid}}(\cR')
    \geq
    \hat{t}_{\mathrm{Sid}}(\cR).
\end{equation*}
Thus, every hypothesis in $\mathcal{D}(\cR)$
is either already contained in $\cR'$ or
also belongs to $\mathcal{D}(\cR')$, establishing
\eqref{sidak_seq_monotonicity}.

We next verify \eqref{sidak_seq_single_step}.
Taking $\cR=[m]\setminus\cH^{u/n}_0$ gives
\begin{equation*}
    \tilde{N}_{\cR}(t)
    =
    \sum_{i\in\cH^{u/n}_0}
    \mathbf{1}\{\tilde{F}_i<t\}.
\end{equation*}
Consequently, $\mathcal{D}(\cR)$ is precisely
the rejection set obtained by applying
AdaFilter-Sid to the subfamily of true
PC nulls $\cH^{u/n}_0$.
Since the assumptions of \thmref{sidak_fwer}$(i)$
continue to hold upon restriction to this
subfamily, we have
\begin{equation*}
    \Pr\left(
        \mathcal{D}([m]\setminus\cH^{u/n}_0)
        \cap\cH^{u/n}_0
        \neq\varnothing
    \right)
    \leq\alpha,
\end{equation*}
establishing \eqref{sidak_seq_single_step}.

Therefore, both conditions of the sequential
rejection principle of \citet{Goeman2010}
are satisfied, and AdaFilter-SeqSid
controls the FWER at level $\alpha$.

We now prove that
$\tilde{\cR}_{\mathrm{SeqBon}}
\subseteq\cR_{\mathrm{SeqSid}}$.
Let $\tilde{\cR}_{\mathrm{Bon}}^{(k)}$ and
$\cR_{\mathrm{Sid}}^{(k)}$ denote the rejection
sets after iteration $k$ of AdaFilter-SeqBon
applied to $\{(\tilde{S}_i,\tilde{F}_i)\}_{i\in[m]}$
and AdaFilter-SeqSid, respectively.

For the former procedure, define the feasible region
\begin{equation*}
    \widetilde{\cT}_{\mathrm{SeqBon}}^{(k)}
    :=
    \left\{
        t\in[0,\alpha]:
        t\tilde{N}_{\tilde{\cR}_{\mathrm{Bon}}^{(k-1)}}(t)
        \leq\alpha
    \right\}
\end{equation*}
and the corresponding rejection threshold
\begin{equation*}
    \tilde{t}_{\mathrm{Bon}}^{(k)}
    :=
    \sup\widetilde{\cT}_{\mathrm{SeqBon}}^{(k)}.
\end{equation*}
If either procedure terminates, we continue
to apply its update rule, which leaves its
rejection set unchanged. Initially,
\begin{equation*}
    \tilde{\cR}_{\mathrm{Bon}}^{(0)}
    =
    \cR_{\mathrm{Sid}}^{(0)}
    =
    \varnothing.
\end{equation*}
Suppose that
\begin{equation*}
    \tilde{\cR}_{\mathrm{Bon}}^{(k-1)}
    \subseteq
    \cR_{\mathrm{Sid}}^{(k-1)}.
\end{equation*}
Then, for every $t\in[0,\alpha]$,
\begin{equation*}
    \tilde{N}_{\cR_{\mathrm{Sid}}^{(k-1)}}(t)
    \leq
    \tilde{N}_{\tilde{\cR}_{\mathrm{Bon}}^{(k-1)}}(t).
\end{equation*}
Consequently, whenever
$t\in\widetilde{\cT}_{\mathrm{SeqBon}}^{(k)}$,
we have
\begin{align*}
    1-(1-t)^{
        \tilde{N}_{\cR_{\mathrm{Sid}}^{(k-1)}}(t)
    }
    &\leq
    t\tilde{N}_{\cR_{\mathrm{Sid}}^{(k-1)}}(t)\\
    &\leq
    t\tilde{N}_{\tilde{\cR}_{\mathrm{Bon}}^{(k-1)}}(t)\\
    &\leq
    \alpha,
\end{align*}
where the first inequality follows from
\eqref{eq:sidak_vs_bon_mult}.
Thus,
\begin{equation*}
    \widetilde{\cT}_{\mathrm{SeqBon}}^{(k)}
    \subseteq
    \cT_{\mathrm{SeqSid}}^{(k)},
\end{equation*}
which implies
\begin{equation*}
    \hat{t}_{\mathrm{Sid}}^{(k)}
    \geq
    \tilde{t}_{\mathrm{Bon}}^{(k)}.
\end{equation*}

Every new hypothesis rejected by the
AdaFilter-SeqBon is therefore either
already contained in $\cR_{\mathrm{Sid}}^{(k-1)}$
or is also rejected by the AdaFilter-SeqSid
at iteration $k$. Consequently,
\begin{equation*}
    \tilde{\cR}_{\mathrm{Bon}}^{(k)}
    \subseteq
    \cR_{\mathrm{Sid}}^{(k)}.
\end{equation*}
By induction,
\begin{equation*}
    \tilde{\cR}_{\mathrm{SeqBon}}
    \subseteq
    \cR_{\mathrm{SeqSid}}.
\end{equation*}

\end{proof}

\section{Additional simulations under dependence}
\label{app:additional_simulations_dep}

We conduct additional simulation studies to investigate
the robustness of the FWER procedures compared in
\secref{simulations} under dependence among the base
$p$-values. We extend the two-point normal mixture model
in \secref{sim_mixture} by introducing an autoregressive
correlation structure of order one (AR(1)) across features
within each study, while maintaining independence
between studies.

Specifically, we independently generate
\begin{equation*}
    \theta_{ij}\sim
    0.5\delta_0+0.5\delta_\mu,
    \qquad (i,j)\in[m]\times[n],
\end{equation*}
and, conditional on the underlying effects, generate
\begin{equation*}
    (Z_{1j},\dots,Z_{mj})^\top \mid
    (\theta_{1j},\dots,\theta_{mj})
    \sim
    \mathrm{N}_m(
    (\theta_{1j},\dots,\theta_{mj})^\top,
    \boldsymbol{\Sigma}_{\rho}),
    \qquad j\in[n],
\end{equation*}
independently across studies, where
$\boldsymbol{\Sigma}_{\rho}$ is an $m\times m$
covariance matrix with elements
\begin{equation*}
    (\boldsymbol{\Sigma}_{\rho})_{ii'}
    =
    \rho^{|i-i'|},
    \qquad i,i'\in[m].
\end{equation*}
We consider $\rho\in\{-0.8,0.8\}$ to examine
strong negative and positive lag-one correlations,
respectively. The two-sided base $p$-values are
computed as $P_{ij}=2\Phi(-|Z_{ij}|)$.

\begin{figure}[t]
    \centering
    \includegraphics[width=0.9\linewidth]
    {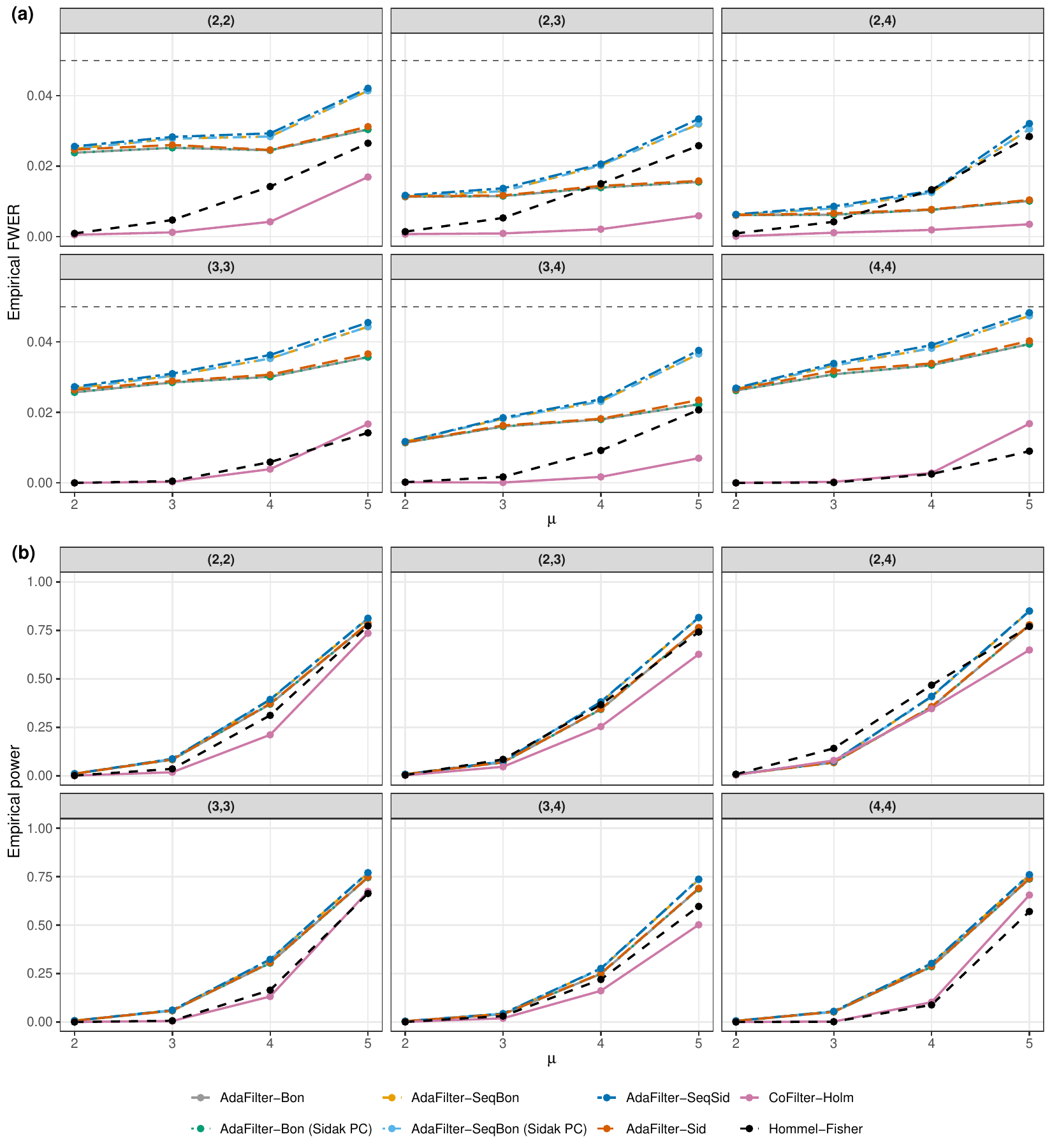}
    \caption{Empirical FWER (panel (a)) and power
    (panel (b)) of the eight competing procedures
    under AR(1) dependence with $\rho=-0.8$.
    Each panel contains six $(u,n)$ configurations,
    with signal strength $\mu\in\{2,3,4,5\}$.
    The horizontal dashed line in panel (a)
    indicates the nominal FWER level $\alpha=0.05$.
    All results are based on $10,000$
    Monte Carlo repetitions.}
    \label{fig:ar1_negative}
\end{figure}

\begin{figure}[t]
    \centering
    \includegraphics[width=0.9\linewidth]
    {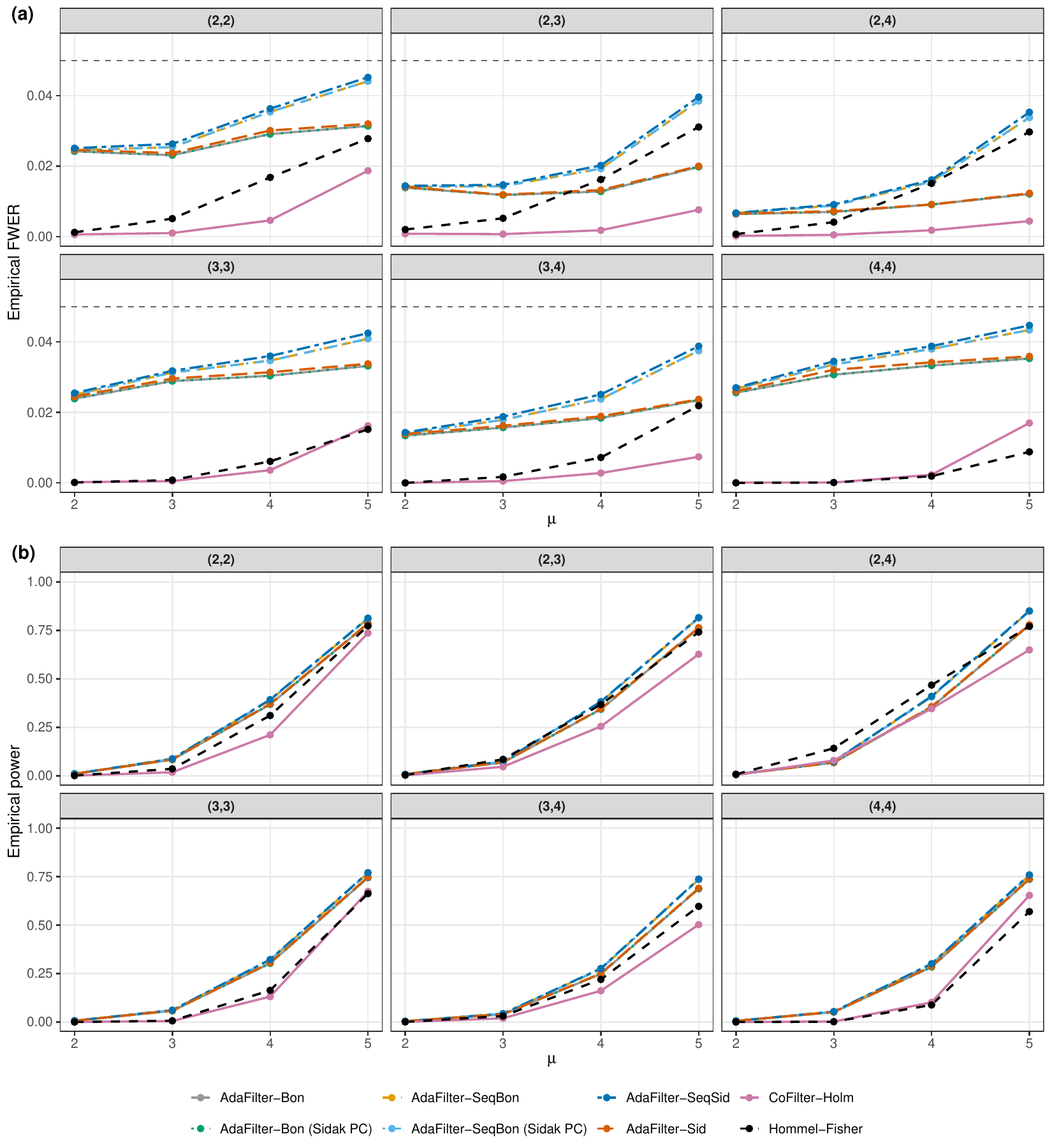}
    \caption{Empirical FWER (panel (a)) and power
    (panel (b)) of the eight competing procedures
    under AR(1) dependence with $\rho=0.8$.
    Each panel contains six $(u,n)$ configurations,
    with signal strength $\mu\in\{2,3,4,5\}$.
    The horizontal dashed line in panel (a)
    indicates the nominal FWER level $\alpha=0.05$.
    All results are based on $10,000$
    Monte Carlo repetitions.}
    \label{fig:ar1_positive}
\end{figure}

\begin{figure}[t]
    \centering
    \includegraphics[width=0.9\linewidth]
    {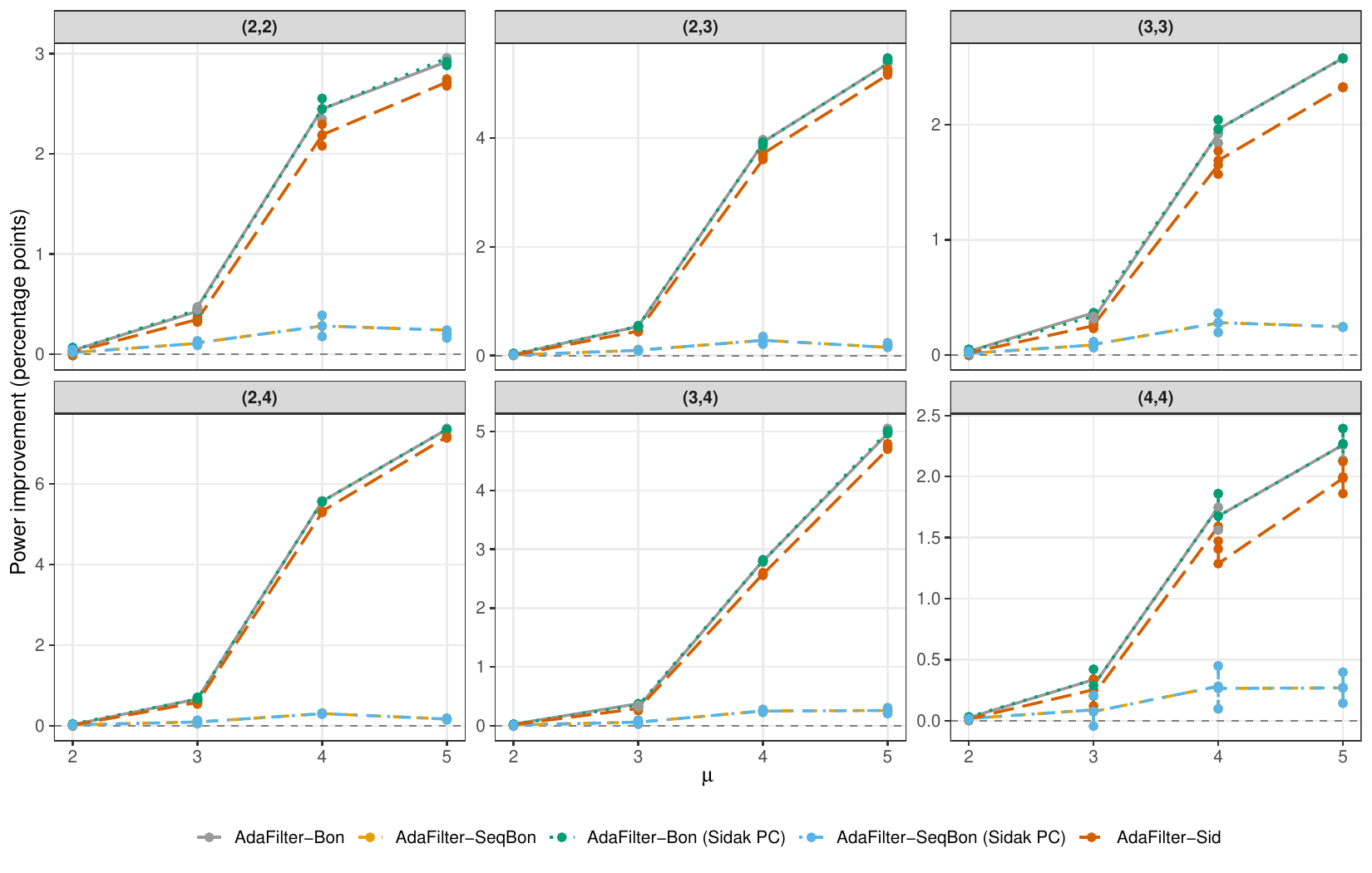}
    \caption{Empirical power improvements (in percentage
    points) of AdaFilter-SeqSid over the other five
    AdaFilter variants under negative AR(1) dependence
    with $\rho=-0.8$, based on $10{,}000$ Monte Carlo
    repetitions. Each panel corresponds to a different
    $(u,n)$ configuration, with signal strength
    $\mu\in\{2,3,4,5\}$. The vertical axes have different
    scales across panels to facilitate comparison of
    the power improvements.}
    \label{fig:ar1_negative_gain}
\end{figure}

\begin{figure}[t]
    \centering
    \includegraphics[width=0.9\linewidth]
    {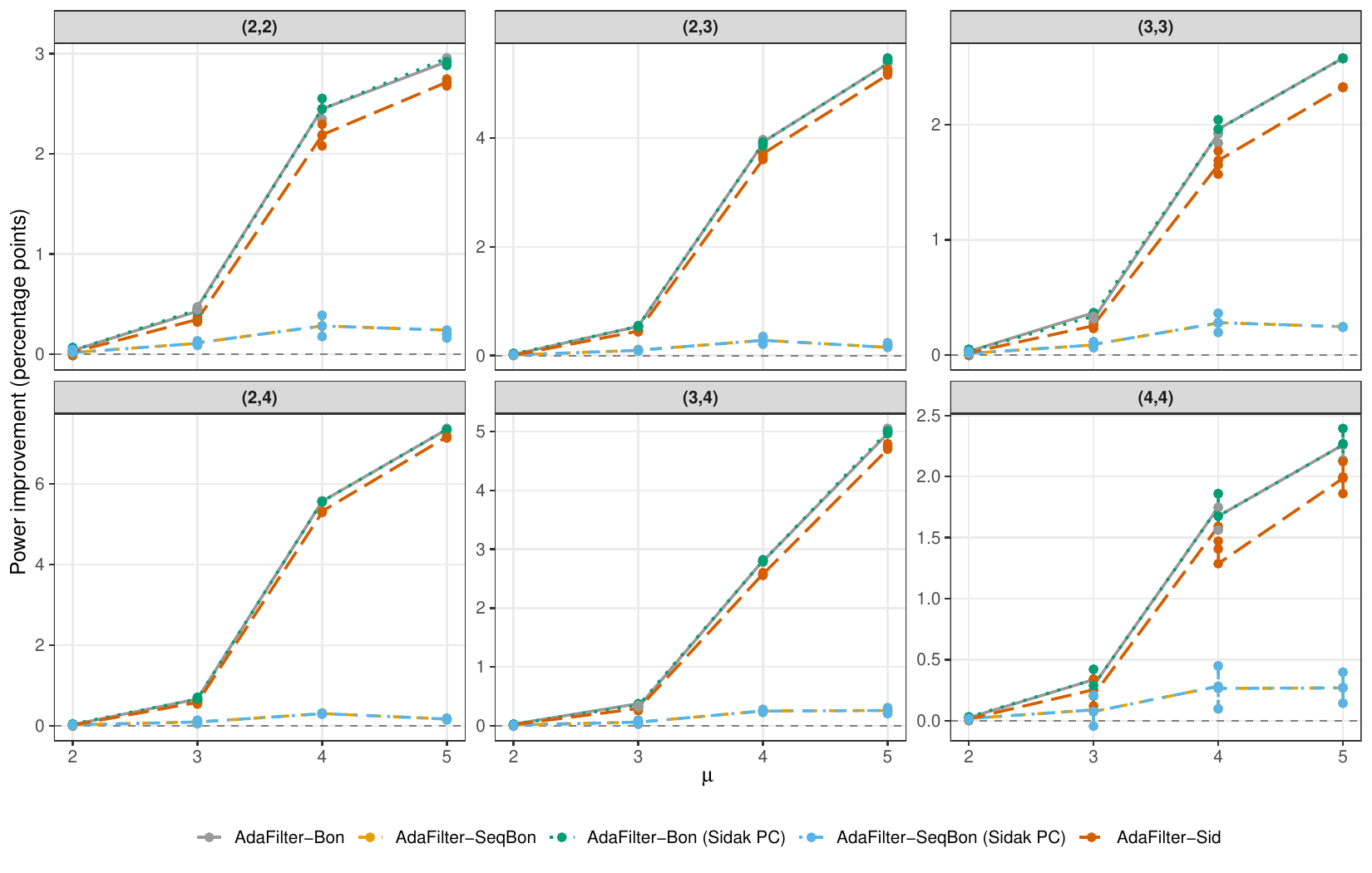}
    \caption{Empirical power improvements (in percentage
    points) of AdaFilter-SeqSid over the other five
    AdaFilter variants under positive AR(1) dependence
    with $\rho=0.8$, based on $10{,}000$ Monte Carlo
    repetitions. Each panel corresponds to a different
    $(u,n)$ configuration, with signal strength
    $\mu\in\{2,3,4,5\}$. The vertical axes have different
    scales across panels to facilitate comparison of
    the power improvements.}
    \label{fig:ar1_positive_gain}
\end{figure}

All other simulation settings remain unchanged
from \secref{sim_mixture}. In particular, we consider
$m=500$ features,
$\mu\in\{2,3,4,5\}$, and the six configurations
\begin{equation*}
    (u,n)\in
    \{(2,2),(2,3),(3,3),(2,4),(3,4),(4,4)\}.
\end{equation*}
For each simulation setting, empirical power and
FWER are estimated from $10,000$ independent
Monte Carlo repetitions at the nominal level
$\alpha=0.05$. The results for negative and positive
lag-one correlations are presented in
\figref{ar1_negative} and
\figref{ar1_positive}, respectively. In each figure,
panel (a) reports the empirical FWER, whereas
panel (b) reports the empirical power. To better
visualise the differences among the AdaFilter variants,
we also report the empirical power improvements of
AdaFilter-SeqSid over the other five AdaFilter variants,
expressed in percentage points as defined in
\secref{sim_mixture}. The corresponding results
are presented in \figref{ar1_negative_gain} and
\figref{ar1_positive_gain} for negative and positive
lag-one correlations, respectively.

Although the rejection-set dominance relationships
between the AdaFilter variants continue to hold,
the dependence structures considered here fall
outside the mutual-independence assumptions underlying
their finite-sample FWER guarantees.
Consequently, the results should be interpreted as
an empirical assessment of robustness rather than
an extension of the theoretical FWER guarantees.
The results yield the following observations:

\begin{itemize}
    \item \textbf{FWER control.}
    Figures~\ref{fig:ar1_negative} and
    \ref{fig:ar1_positive} show that all eight
    procedures maintain empirical FWER below the
    nominal level $\alpha=0.05$ across all simulation
    settings, under both negative and positive
    lag-one correlations. This provides empirical
    evidence of robustness to violations of the
    mutual independence assumption.

    \item \textbf{Empirical power.}
    Empirical power generally increases with signal
    strength $\mu$ under both dependence structures.
    Among the six AdaFilter variants,
    AdaFilter-SeqSid consistently achieves the
    highest power, in agreement with its theoretical
    rejection-set dominance. The sequential
    variants also generally outperform their
    non-sequential counterparts.

    \item \textbf{Positive versus negative dependence.}
    The empirical FWER and power patterns are
    broadly similar for $\rho=-0.8$ and $\rho=0.8$,
    suggesting that the competing procedures are
    relatively insensitive to the sign of the
    lag-one correlation under the considered AR(1)
    model.

    \item \textbf{Power improvements.}
    \figsref{ar1_negative_gain} and
    \ref{fig:ar1_positive_gain} show that the power
    improvements of AdaFilter-SeqSid over the other
    five AdaFilter variants remain nonnegative,
    as guaranteed by the rejection-set dominance
    relationships, which hold regardless of the
    dependence structure.
    The improvements generally become more pronounced
    as $\mu$ increases, particularly relative to the
    non-sequential variants, reaching approximately
    $7.36$ percentage points for $(u,n)=(2,4)$
    and $\mu=5$ under both $\rho=-0.8$ and $\rho=0.8$.
    In contrast, the improvements over the other
    sequential AdaFilter variants remain relatively
    modest, never exceeding $0.31$ percentage points.
    These findings closely resemble those under
    independence in \secref{sim_mixture}, suggesting
    that the power advantages of the proposed
    procedures persist under the dependence
    structures considered.
\end{itemize}

\clearpage

\section*{Data and Code Availability}
The R code for reproducing the simulation results and real data analysis is available at \url{https://github.com/ninhtran02/AFSid}.

\bibliographystyle{anzjs}
\bibliography{ejs-reference}
\end{document}